\documentstyle[12pt,aasms]{article}

\received{July 22, 1996}
\accepted{October 17, 1996}
\lefthead{Calzetti}
\righthead{Star Formation in Starburst Galaxies}
\slugcomment{To appear in the Astronomical Journal}

\begin{document}

\title{Reddening and Star Formation in Starburst Galaxies.}

\author{Daniela Calzetti\altaffilmark{1}}
\affil{Space Telescope Science Institute, 3700 San Martin Dr., 
   Baltimore, MD 21218, USA}

\altaffiltext{1}{Visiting Astronomer, Kitt Peak National Observatory, operated
by the Association of Universities for Research in Astronomy, Inc. under
contract with the National Science Foundation.} 

\begin{abstract}

The reddening properties and the star formation history of a sample of
19 starburst galaxies are investigated using multiwavelength
spectroscopy and infrared broad band photometry. New photometric data
in the J, H, and K bands of the central starburst regions are
supplemented with previously published spectra, covering the
wavelength range 0.12-2.2~$\mu$m.  In the infrared, the reddening
value derived for the stellar continuum is in agreement with that of
the ionized gas, but the two values diverge at shorter wavelengths; in
the UV, the mean optical depth of the dust in front of the stars is
smaller, being only 60\%, than the optical depth of the dust in front
of the nebular gas. This difference can be better understood if the
UV-bright stellar population and the ionized gas are not co-spatial. A
model of foreground clumpy dust, with different covering factors for
the gas and the stars, is proposed to account for the difference in
reddening. A ``template starburst spectrum'', derived by combining the
reddening-corrected UV, optical, and infrared data of all the galaxies
in the sample, is used to investigate the star formation history of
the galaxies. Spectral synthesis models indicate that the observed UV
emission can be attributed to a stellar population which is undergoing
active star formation at a constant rate since
$\sim$2$\times$10$^7$~yr, in agreement with the supernova rates
derived from the [FeII] emission line in the infrared. At least two,
and probably more, intermediate age populations
(age$<$2$\times$10$^9$~yr) contribute to the optical and infrared
emission, while populations older than $\sim$2$\times$10$^9$~yr do not
contribute significantly to the template. The stellar composition of
the template spectrum suggests episodic star formation over the last
10$^9$~yr, with star formation rates as large as or larger than the
present rates. The synthetic stellar populations are generated
according to an Initial Mass Function (IMF) with Salpeter slope
($\alpha$=2.35) in the mass range 0.1--100~M$_{\odot}$, and reproduce
a number of observational constraints, such as the spectral energy
distribution of the template spectrum, the equivalent width of the
atomic hydrogen emission lines, and the mass-to-light ratios; the
data, then, do not provide indication for a high-mass-star~truncated
or a low-mass-star~deficient IMF in starburst galaxies.
\end{abstract}


\section{Introduction}

The study of the history of stellar populations plays a crucial role
in understanding the evolution of galaxies, and, in this framework,
starbursts may occupy a key position. In a starburst galaxy, the
radiation output is dominated by the emission from a large number of
recently formed, hot, massive stars. In some cases, the rate of star
formation is high enough that, if it remains constant, the galaxy will
be depleted of its gas content in much less than a Hubble time. The
role of starbursts within the galaxy evolution scenario is still
debated: are bursts of star formation the preferred channel for
forming new stellar populations or are they just sporadic events in
the lifetime of some galaxies? Did the first generation of stars in a
primeval galaxy form from a widespread burst of star formation over a
short timescale or was it created through a slow and mostly quiescent
process?

In order to answer these questions, we need first to understand how
starbursts evolve in local galaxies. Star formation episodes show a
large variability in duration, ranging from 10$^7$--10$^8$~yr (Rieke
et al. 1985) to more than 10$^9$~yr (Hunter \& Gallagher 1986).  The
lifespan of a star formation event is only loosely constrained by the
gas depletion timescale. The mechanisms for generating inflows of gas
to fuel the star forming region are not completely understood yet,
although models are being investigated (e.g. Shlosman 1992).  The
total gas mass in a galaxy is difficult to determine, since both the
atomic and molecular gas content need to be measured, but
observational techniques to derive the molecular gas mass may be
inadequate (e.g. Allen 1996). The gas mass available for star
formation may be increased by inputs from interacting companions. Once
triggered, the evolution of a starburst region depends on the details
of its interaction with the galaxy environment. For instance, bursts
of star formation can propagate across a galaxy, increasing the
duration of the event (e.g., Beck et al. 1996). High resolution
imaging reveals that bursts of star formation break down into distinct
knots of OB associations, often characterized by different
evolutionary stages (Meurer et al. 1995, Conti \& Vacca
1994). Finally, star formation processes can be depressed by gas
outflows which may develop into superwinds and deplete a region of
gas, and, therefore, of fuel (Heckman et al. 1990, De~Young \& Heckman
1994). It follows that feedback mechanisms from the host galaxy
influence the evolution and the duration of a burst of star formation,
and this is reflected into the stellar populations generated by the
star-forming episode.

The universality of the stellar Initial Mass Function (IMF) in
starbursts is also an open issue. A dependence of the IMF on
environment may produce dramatically different results for the star
formation history of the galaxy.  Increasing evidence suggests,
however, that the slope of the IMF in OB associations is fairly
independent of the gas metallicity and close to the Salpeter value
($\alpha$=2.35) in the high-mass~stars regime (Massey et al. 1995,
Moffat 1996, Stasinska \& Leitherer 1996). The upper and lower mass
limits for the IMF in starburst galaxies are still strongly
debated. Stars as massive as 85--100~M$_{\odot}$ are found in the
Galaxy, in the Magellanic Clouds, and in nearby resolved starbursts,
suggesting upper mass limits independent of the environment (Kudritzki
et al. 1991, Massey et al. 1995, Moffat 1996). Infrared studies of
unresolved starburst galaxies point, however, at upper mass limits as
low as 30~M$_{\odot}$ for the IMF, using either the comparison of the
observed spectra with models, as in the case of the luminous infrared
galaxies (Goldader et al. 1996), or the analysis of the He/H line
emission ratios (Doyon, Puxley, \& Joseph 1992, see however Shields
1993). Even more difficult is the determination of the low~mass~star
limit, which cannot be directly observed, because the energy output
from a starburst galaxy is dominated by the high~mass stars produced
in the burst. Values larger than 1~M$_{\odot}$ for the low~mass limit
of the IMF in starburst galaxies have been suggested on the basis of
the observed mass-to-light ratios (Rieke et al. 1980, 1993, Bernl\"ohr
1993). These conclusions, however, depend on the adopted IMF (Rieke et
al. 1993) and on the reddening correction applied to the
luminosity. Revisions of the latter indicate smaller values of the
low~mass~star limit, at least for the famous case of M~82 (Lester et
al. 1990, Satyapal et al. 1995). The analysis of a relatively large
sample of galaxies shows that a low-mass-star~deficient IMF is,
indeed, not universal among starbursts (Devereux 1989).

Corrections for the effects of dust reddening are crucial when
multiwavelength observations are used to derive intrinsic quantities
for a galaxy. Dust selectively extincts the radiation, with blue
wavelengths being more affected than red wavelengths. As an added
complication, limitations in angular resolution usually produce
observations of unresolved galactic regions, and the {\it effective}
reddening of the emerging radiation is determined by the details of
the dust-stars-gas geometry (Natta \& Panagia 1984, Disney, Davies \&
Phillips 1989, Witt, Thronson \& Capuano 1992, Byun, Freeman, \&
Kylafis 1994, Giovanelli et al. 1995, Xu \& Buat 1995, Shields \&
Kennicutt 1995, Witt \& Gordon 1996, Boisse\'e \& Thoraval 1996). Dust
obscuration in moderately reddened starburst galaxies can be well
represented by a (clumpy) dust distribution foreground to the star
forming region(s) (Calzetti, Kinney, \& Storchi-Bergmann 1994, 1996,
Kinney et al. 1994, Calzetti et al. 1995, Meurer et
al. 1995). However, the reddening of the stellar continuum in these
galaxies is generally lower than the reddening of the ionized gas
(e.g.  Fanelli, O'Connell \& Thuan 1988, Calzetti et al. 1994, CKS94
hereafter, Lan\c{c}on, Rocca-Volmerange \& Thuan 1996, Mayya \& Prabhu
1996), suggesting that the covering factor of the dust in front of the
stars is lower than that in front of the gas. Generally, the effects
of the dust obscuration are only partially explained by the foreground
distribution. The observed UV and optical radiation accounts only for
70\% of all the radiation contributing to the far~infrared (IRAS)
luminosity (Calzetti et al. 1995), implying that a fraction of the
stars is embedded in optically thick dust clouds.

The study of the star formation histories of starburst galaxies helps
to gain insight into the starburst phenomenon. New photometric data at
J, H, and K are presented and discussed for a sample of 24 galaxies,
and are supplemented with published UV, optical and near infrared
spectra. These data are used to investigate the reddening properties
of the stellar continuum in the entire wavelength range UV-to-IR, and
to derive the stellar content of starburst regions. The investigation
on the dust obscuration of the stellar continuum presented here
(section~4) is an extension to infrared wavelengths of the study of
CKS94; the results on the reddening of the ionized gas of Calzetti et
al. (1996, CKS96 hereafter) will be used. The observations are
presented in sections~2 and 3. Constraints on the recent star
formation history through the Type II supernova rates are derived in
section~5, and the stellar content and star formation history of the
galaxies are discussed in sections~6 and 7. The conclusions are
presented in section~8.

\section{The Data}

\subsection{Sample Description}

The 24 galaxies listed in Table~1 include 19 starbursts, 1 LINER, and
4 Seyfert~2s, selected from the {\it Atlas of Ultraviolet Spectra of
Star-Forming Galaxies} of Kinney et al. (1993). A galaxy had to have a
well detected UV spectrum in the IUE archive in order to be included
in the sample, which automatically excluded heavily reddened
objects. In addition to this basic selection criterion, optical
spectra obtained in apertures which closely match that of IUE had to
be available (see section~2.3). For this sample, broad-band images at
J, H, and K, and, for some of the galaxies, infrared (IR) spectra were
obtained (see CKS96 for the IR spectroscopy).

The 24 galaxies are at low redshifts, with distances between 2.5~Mpc (NGC~1569)
and 300~Mpc (Mrk~357), with median value $\sim$60~Mpc
(H$_{\circ}$=50~km/s/Mpc). The multiwavelength data are obtained using
observational apertures which are closely matched in size, so that the same
region is sampled at all wavelengths in each galaxy. The apertures subtend
physical regions of about 4.5~kpc at the typical distance of the galaxies,
implying that a fairly large portion of the galaxy population is sampled. The
19 starburst galaxies span a range of morphological types, from Sab to
Irregulars, usually with disturbed morphologies. The oxygen abundance ranges
typically from 1/4 to twice solar (see Table~2), although abundances as low as
1/10 solar have been derived for NGC~4861 (Barth et al. 1994). A detailed
description of the data used in this work is given in the following two
sections. 

\subsection{The Infrared  Images}

The IR images were obtained with the Simultaneous Quad-color Infrared
Imaging Device (SQIID) at the KPNO 1.3-m telescope during two separate
runs in April 1994 and October 1994. The SQIID has four detectors
which are exposed simultaneously to image a field in the J, H, K, and
L' bands. Each detector has 256$\times$256 pixels, with a scale of
1.395$^{\prime\prime}$/pix, 1.366$^{\prime\prime}$/pix, and
1.361$^{\prime\prime}$/pix at J, H, and K, respectively. The
overlapping area between the detectors is 245$\times$245 pixels (Ellis
et al. 1992).

The targets were typically observed with multiple exposures
of 3 minutes each, to account for the fluctuations of the sky
background. Total integration times ranged from 15 minutes for the
brightest galaxies to 36 minutes for the faintest galaxies. NGC~1068
was observed with multiple exposures of 5 seconds each, with total
integration time of 30 seconds. Sky frames were obtained by moving the
targets across the SQIID 5.5$^{\prime}$ field of view.

The data were background subtracted using DIMSUM\footnote[1]{DIMSUM is
the Deep Infrared Mosaicing Software package developed by
P. Eisenhardt, M. Dickinson, A. Stanford, and J. Ward, and is
available upon request to prme@kromos.jpl.nasa.gov or med@stsci.edu},
which masks the objects and rescales the images before median
filtering to create the sky frames.  Flat-fielding and co-addition of
the images was obtained using the SQIID data reduction
package\footnote[2]{The SQIID data reduction package has been
developed by M.~M. Merrill and J.~W. MacKenty and is available upon
request to merrill@noao.edu}.

At least one blue (J$-$K$<$0.1) and one red (J$-$K$\sim$0.9) standard
stars from the list of Elias et al. (1982) were observed each night
for flux calibration and color corrections. The blue standards were
followed through the night to provide airmass corrections. The loss in
flux due to a unit change in airmass is 0.175 mag at J and K, and
0.125 mag at H.

Cross-calibration of the standard stars shows variations of about 6\%
in the absolute K band photometry during the April run and 7\% in the
same band during the September run. The uncertainties are smaller in
the more sensitive J and H bands.  The variations of the observed
colors relative to the expected ones for the standard stars are
$\Delta (J-H)< 3$\% and $\Delta (H-K)< 2$\% during the April run, and
$\Delta (J-H)<4$\% and $\Delta (H-K)< 3$\% during the September
run. The variations are small, likely because of the simultaneous
exposure of the J, H, and K detectors.

The observed galaxies are listed in Table~1, together with the
activity type, the redshift, the K-band magnitude, the J$-$H color,
and the H$-$K color, measured in an aperture of 10$^{\prime\prime}$
diameter centered on the starburst. All quantities have been corrected
for the foreground extinction from our Galaxy (Burstein \& Heiles
1982). The present sample has some overlap with the sample of
Markarian galaxies analyzed by Balzano \& Weedman (1981), who obtained
absolute photometry at J, H, and K in an aperture comparable to the one
used here. The 7 galaxies in common between the two samples are:
IC~1586 (Mrk~347), NGC~1614 (Mrk~617), NGC~4194 (Mrk~201), NGC~4385
(Mrk~52), NGC~6052 (Mrk297), NGC~6217, NGC~7714 (Mrk~538). Differences
$\Delta K < 0.09$ and $\Delta J < 0.12$ are observed in the absolute
photometry of the common galaxies; typical differences in the colors
are less than $\Delta (J-H)\simeq 0.07$ and $\Delta (H-K)\simeq
0.05$. We adopt the differences as 1~$\sigma$ uncertainties on our
measurements. The differences with previous results are larger for the
absolute photometry than for the colors, and can be attributed to the
different filter bandpasses used in the two works, to the different
calibration stars, and to atmospheric transparency variations during
our observations.

\subsection{The Ultraviolet, Optical and Infrared Spectra}

The ultraviolet (UV) and optical spectra of the 19 starburst galaxies
listed in Table~1 are presented in Kinney et al. (1993), McQuade,
Calzetti, \& Kinney (1995), and Storchi-Bergmann, Kinney, \& Challis
(1995). The UV (IUE) and optical spectra cover the spectral range
0.12--0.80~$\mu$m. Quantities measured from those spectra are given in
Table~2 (see also CKS94). For a subsample of 16 galaxies near-infrared
(NIR) spectra were obtained (CKS96) and the data relevant to the
present work are reported in Tables~3 and 4. The NIR spectra cover the
wavelength range 1.236--1.340~$\mu$m in the J band and
2.113--2.250~$\mu$m in the K band. The optical and NIR observational
apertures were closely matched to the IUE aperture
(10$^{\prime\prime}\times$20$^{\prime\prime}$), in order to compare
similar regions at different wavelengths in each galaxy. Some of the
NIR spectra are shown in Figure~1, and two examples of multiwavelength
spectra are given in Figure~2.

For the 19 starburst galaxies in the sample, Table~2 reports: the
intrinsic color excess from the Balmer line ratio H$\alpha$/H$\beta$
(CKS94), the oxygen abundance $log(O/H)+12$ (Storchi-Bergmann,
Calzetti, \& Kinney 1994), the H$\alpha$ equivalent width (EW), the UV
flux density F(0.16) at 0.16~$\mu$m, the K-band flux density F(2.22),
and the NIR spectral index $\beta_{im}$. All quantities are corrected
for the foreground Galactic reddening. The UV flux density is the mean
value derived from the IUE spectra in the range
0.125--0.195~$\mu$m. The K-band flux density F(2.22) from the images 
is measured in a 10$^{\prime\prime}\times$20$^{\prime\prime}$ window
centered on the starburst region, to match the UV aperture.  The IR
magnitudes are converted into flux densities using the zero-points of
Campins et al. (1985). The NIR spectral index $\beta_{im}$ is derived
from the flux densities at J, H, and K, assuming that the IR spectral
energy distribution can be described by the power law
$F(\lambda)\propto \lambda^{\beta_{im}}$. The J, H, and K photometric
flux densities used in the fit are also extracted from a
10$^{\prime\prime}\times$20$^{\prime\prime}$ window, with the same
orientation of the NIR spectra (E--W), to avoid aperture mismatches
with the spectroscopic data.

Tables~3 and 4 contain the data from the NIR spectroscopy: the flux
and EW of the hydrogen emission lines Pa$\beta$(1.282~$\mu$m) and
Br$\gamma$(2.166~$\mu$m), the flux of the [FeII](1.257~$\mu$m) and the
H$_2$~$v=1-0S(1)$~(2.121~$\mu$m) emission lines, the intrinsic galaxy
reddening from the Balmer/Brackett line ratio H$\beta$/Br$\gamma$, the
K-band flux density F(2.18), the NIR spectral index $\beta_{sp}$, the
reddening corrected ratio [FeII](1.257~$\mu$m)/H$_2$(2.121~$\mu$m),
and the ratio between the two H$_2$ transitions
$v=1-0$S(0)~(2.223~$\mu$m) and $v=1-0$S(1)~(2.121~$\mu$m).  For the
latter, only one measurement and five upper limits could be
determined. For the other galaxies, the redshift is high enough that
the H$_2$ line at 2.223~$\mu$m is outside the spectral range used
here. The upper limits on H$_2$(2.223~$\mu$m) are calculated from the
flux of the largest noise spike in the wavelength region where the
line is expected to be located. The spectral index $\beta_{sp}$ is
derived via a simultaneous $\chi^2$ fit of the J and K spectra with a
power law $F(\lambda)\propto \lambda^{\beta_{sp}}$, after removal of
the emission lines. The flux density F(2.18) is the mean value derived
from the wavelength range 2.115-2.250~$\mu$m.  The Pa$\beta$ and
Br$\gamma$ line intensities, and the reddening values from
H$\beta$/Br$\gamma$ appeared in CKS96 for 13 of the 16 galaxies in
Table~3; those values are listed here for convenience, since they will
be frequently used in the following sections.  The comparison of the
Pa$\beta$ and Br$\gamma$ line fluxes measured here with values
published in the literature is discussed in CKS96. For the [FeII] and
H$_2$(2.121~$\mu$m) emission lines, only three galaxies are in common
between this and other samples: the H$_2$ line fluxes of NGC1614 and
NGC6090 are given in Goldader et al. (1995); the [FeII] line flux of
NGC1614 in Puxley et al. (1994); and the H$_2$ and [FeII] line fluxes
of NGC7714 in Moorwood \& Oliva (1988). The aperture sizes used by
these authors are 4 to 5 times smaller than those used here. Their
line measurements give fluxes which are 2 to 5 times smaller than
ours, pointing at the extended nature of the [FeII] and H$_2$ emission
in these galaxies.

\subsection{The Infrared Images-Spectra Comparison}

The K-band flux densities from the images are systematically 10--15\%
lower than the analogous flux densities from the spectra. The spectra
cover only $\sim$0.1~$\mu$m in wavelength range, about 1/6 of the
typical K bandpass used for the images. The transformation from one to
the other, therefore, carries uncertainties due to the unknown
behaviour of the galaxies energy distribution in the wavelength range
not covered by the spectra. This, coupled with the fact that
variations of the atmospheric transparency were larger during the
imaging observations than during the spectroscopic observations, may
ultimately account for the mismatch.

The NIR spectral index derived from the spectra, $\beta_{sp}$ must
have the same value of the index from the images, $\beta_{im}$, if the
relative calibration of different wavebands has been done
correctly. Figure~3 shows that this is the case for 14 of the 16
galaxies in Table~4. The two exceptions are IC~1586 and Mrk~542, whose
spectra were obtained in poor weather conditions. An uncertainty of
less than 30\% in the calibration of the J spectra relative to the K
spectra produces a variation $\delta \beta_{sp} = 0.5$. In the next
sections, the NIR colors associated with the two galaxies will be
those from the imaging, since the J, H, and K images were obtained
simultaneously, and are thus almost independent of the sky conditions.

\section{Infrared Colors}

The starburst, LINER, and Sy2 galaxies in our sample occupy the
typical loci expected for these classes of galaxies in the
(J$-$H)--(H$-$K) diagram (see Figure~4; e.g. Carico et al. 1986). Only
NGC~4861 and NGC~1068 deviate from the general trend. The J$-$H color
of NGC~4861 is bluer than the other starbursts and is due to the
Pa$\beta$ emission line's contribution to the J-band magnitude
(Table~3). The Pa$\beta$ EW of 200~\AA\ corresponds to about 9\% of
the J-band flux; after the line removal, the galaxy's IR color becomes
J$-$H=0.47, closer to the expected value for the stellar continua of
bursts of star formation.

The Seyfert 2 and LINER galaxies in our sample, except NGC~1068, are
indistinguishable from the starbursts in the (J$-$H)--(H$-$K) plane
(e.g.  Carico et al. 1990). Although the intrinsic colors for Seyfert
nuclei are expected to be J$-$H$\sim$0.95 and H$-$K$\sim$1.05 (Glass
1981), the 10$^{\prime\prime}$ aperture subtends galaxy regions of
$\sim$2--4~kpc in size, and the contribution to the IR flux from the
host galaxy and/or the circumnuclear burst of star formation is
significative (e.g. Heckman et al. 1995a).  The NIR colors of NGC~1068
are much redder than the other 3 Sy~2's, as already known from
previous investigations (e.g. Scoville et al.  1988). This galaxy is
closer to our Galaxy than the other Sy2's by a factor 2 to 4. The
physical region sampled by the 10$^{\prime\prime}$ aperture has a
linear scale of about 1~kpc, smaller than the inner radius of the
circumnuclear starburst ring (about 1.5~kpc, Heckman et
al. 1995a). The IR colors of NGC~1068 become bluer for increasing
aperture size (Figure~4, cf. Scoville et al. 1988), as expected if the
contribution from the stars of the host galaxy increases.  The colors
of an annulus with inner and outer radii of 15$^{\prime\prime}$ and
30$^{\prime\prime}$, respectively, embracing the circumnuclear burst
of star formation (Heckman et al. 1995a), are consistent with those of
the starburst galaxies. The H$-$K color of the nucleus of NGC~1068 (in
the 5$^{\prime\prime}$ aperture) is redder than the expected value for
Seyferts.  If the intrinsic color of the non-thermal nuclear source is
H$-$K=1.05, and if hot dust emission in the K-band is the cause for
the red H$-$K color (Carico et al. 1990), then the hot dust provides
about 50\% of the flux at 2.2~$\mu$m.

\section{Reddening}

\subsection{The Reddening of the Stellar Continuum}

The properties of the starburst galaxies in the sample cannot be
properly discussed without investigating the impact of the dust
obscuration on the measured fluxes. The analysis of the dust reddening
in external galaxies is usually complicated by the potentially complex
geometries which characterize the distribution of stars, ionized gas,
and dust. Even in the presence of fairly large amounts of dust, mixed
distributions of the various components can mimick a grey obscuration,
and give the (incorrect) impression that the reddening in the galaxy
is negligible (e.g., Witt, Thronson, \& Capuano 1992).  In addition,
the obscuration suffered by the stellar continuum may be different
from the obscuration of the ionized gas, since stars and gas may
occupy different locations within the galaxy, and the dust may have
different covering factors for stars and gas. Various authors (e.g.,
Fanelli et al. 1988, CKS94, Lan\c{c}on et al. 1996) have reported that
the emission from stars is typically less obscured than the emission
from the ionized gas. However, CKS94 showed that the reddening of the
UV continuum emission (i.e., massive stars) correlates with the
reddening of the gas.  In this and the next section we use the
infrared observations presented in section~2 to pin down the effects
of obscuration on the stellar continuum of the sample galaxies, and to 
derive a description for the geometrical distribution of the dust. For
the reddening of the nebular gas, the results by CKS96 are adopted.

From the analysis of the hydrogen emission line intensities at optical
and infrared wavelengths, CKS96 conclude that the geometry of the dust
obscuring the ionized gas can be described by a foreground clumpy
distribution, and the reddening is compatible with a Galactic-like
extinction curve.  Although the solution for the geometry is not
unique (see Puxley \& Brand for the case of NGC1614), it represents a
simple description which accounts for the observed reddening values of
the gas in the optical$+$IR wavelength range (up to Br$\gamma$).  The
same study excludes pure internal dust as a viable geometrical
description.

The dust obscuration becomes progressively less important towards IR
wavelengths; the Galactic extinction is about 10 times smaller at
2.2~$\mu$m than at 0.55~$\mu$m. The effects of dust reddening on the
stellar continuum can then be investigated using multiwavelength
observations. If the shape of the galaxy continuum spectrum is determined
by the overall effect of the dust, rather than by the details of  the 
intrinsic stellar population, the  
variations of the continuum strength can be interpreted as due to dust
reddening. The ratio of the flux density 
at $\lambda_X$ to the K-band flux density, $\log
F(\lambda_X)/F(K)$, can be studied as a function of the color
excess E(B$-$V)$_{H_a/H_b}$, derived from a pair of hydrogen emission lines
 H$_a$ and H$_b$, to gain insight in the geometrical
distribution of the dust associated with the starburst region.  The
standard definition of the color excess (see, e.g. CKS96):
$$E(B-V)_{H_a/H_b}= \displaystyle{\log (R_{obs}/R_{int}) \over
\displaystyle{ 0.4 [k(\lambda_a)-k(\lambda_b)]}},\eqno(1)$$ 
will be used in the following derivation.  R$_{int}$ and R$_{obs}$ are the
intrinsic and observed hydrogen line ratios, respectively, and
k($\lambda$) is the extinction curve, measured at the wavelength of
the emission line.  The curve k($\lambda$) is defined as the
total-to-selective extinction of the gas, A($\lambda$)/E(B$-$V), where
A($\lambda$) is the attenuation in magnitudes. Both the extinction curves 
of the Galactic diffuse ISM and of the 30~Doradus region in the Large 
Magellanic Cloud are used in the analysis. At optical wavelengths, 
the Galactic curve is parametrized according to Seaton (1979) and
the 30~Dor curve according to Fitzpatrick (1986). 
 In the IR, the parametrization of Landini et al. (1984) is
adopted for both the extinction curves. For the color excess, both the 
standard Balmer ratio H$\alpha$/H$\beta$ and the ratio H$\beta$/Br$\gamma$ 
are used. The latter has the advantage of a long wavelength 
baseline, which reduces the effects of measurement uncertainties (CKS96). 
The adopted values for the differential extinction suffered by the 
hydrogen lines are: k(H$\beta$)$-$k(H$\alpha$)=1.17 and 
k(H$\beta$)$-$k(Br$\gamma$)=3.35.

The F($\lambda_X$) flux densities selected for the analysis are
centered at $\lambda_X$=0.16, 0.26, 0.44, 0.55, 0.70, 1.25, and
1.65~$\mu$m. The UV flux densities, F(0.16) and F(0.26), are the mean
values in 0.06~$\mu$m wide emission bands; the optical flux densities
are obtained from 0.09~$\mu$m or 0.1~$\mu$m wide bands, after removal
of the emission lines. The F(1.25) and F(1.65) flux densities are from
the J and H band measurements, respectively (Tables~2 and 4). The
K-band flux densities are used as reference point for the continuum
emission at shorter wavelengths. Examples of stellar continuum ratios
as a function of the color excess measured from the hydrogen
recombination lines are given in Figure~5.

A (mostly) linear correlation can be discerned at all wavelengths, and
the linear fits to the flux density ratios are expressed as:
$$\log [F(\lambda_X)/F(K)] =\ s\ E(B-V)_{H_a/H_b} +\ \log
[F_{\circ}(\lambda_X)/F_{\circ}(K)],\eqno(2)$$ 
where $F_{\circ}(\lambda_X)/F_{\circ}(K)$ is the intrinsic flux density
ratio at the specified waveband $\lambda_X$. For
H$_a$/H$_b$=H$\alpha$/H$\beta$, F(K)=F(2.2), whereas for
H$_a$/H$_b$=H$\beta$/Br$\gamma$, F(K)=F(2.18).  The linear correlation
between the stellar continuum flux ratios and the color excess of the
ionized gas indicates that the reddening of the gas is proportional
that of the stars (CKS94).  Although complex geometrical distributions
of dust, gas, and stars still are not excluded, an `effective'
obscuration curve for the stellar continuum can be defined as:
$$ k'(\lambda_X) - k'(K) = -s/0.4,\eqno(3)$$ 
where 
$$k'(\lambda)={\displaystyle{A'_{star}(\lambda)}\over \displaystyle
E(B-V)_{gas}},\eqno(4)$$ 
with A$^{\prime}_{star}$($\lambda$) the effective attenuation of
the stellar continuum.  The selective obscuration of the stellar
continuum, k$^{\prime}$($\lambda_X$)$-$k$^{\prime}$(K), is then
normalized to the color excess of the gas.

The list of values for k$^{\prime}$ at different wavelengths is
reported in Table~5, together with the intrinsic flux density ratios,
the number of points used in each fit, and the reduced $\chi^2$.  The
continuum flux density ratios are fitted using
E(B$-$V)$_{H\alpha/H\beta}$ as independent variable for the IR images,
and E(B$-$V)$_{H\beta/Br\gamma}$ for the spectra. The lower number of
data available from spectroscopy is usually compensated by tighter
correlations, resulting in smaller values of the reduced $\chi^2$ and
in comparable 1~$\sigma$ uncertainties in the two data sets.  The
galaxies NGC~4861 and IC~214 do not follow the same trend as the
others at UV and optical wavelengths, respectively, and are removed
from the fits. NGC~4861 is in any respect a peculiar galaxy in our
sample; the large EWs of its emission lines suggest an overall stellar
population which is much younger than the average of the other
galaxies in the sample. Both the continuum and the line emission from
NGC~4861 are dominated by the Bright Knot, a giant HII region with an
upper limit on the age of $\sim$4.5~Myr (Barth et al.  1994, see also
section~7).  For IC~214, the deviation can be attributed to the
absolute calibration of the optical spectrum, which appears
underestimated. There are only 8 galaxies in our sample for which
F(0.26) is available, therefore the data point
k$^{\prime}$(0.26)-k$^{\prime}$(K) has the largest uncertainty. The
values of k$^{\prime}$($\lambda$) are shown in Figure~6, together with
the curve obtained by CKS94 at UV and optical wavelengths. CKS94 could
not derive the absolute value for the obscuration curve, and their curve
is here arbitrarily placed at k$^{\prime}$(0.55)-k$^{\prime}$(2.2)=2,
which is approximately the halfway point between the two sets of
values obtained from E(B$-$V)$_{H\alpha/H\beta}$ and from
E(B$-$V)$_{H\beta/Br\gamma}$ (see Table~5). The
k$^{\prime}$($\lambda$) values derived from the
E(B$-$)$_{H\beta/Br\gamma}$ correlations are systematically higher
than the values derived from the E(B$-$)$_{H\alpha/H\beta}$
correlations. The two sets of data do not appear to be drawn from two
different populations; however, the $H\beta/Br\gamma$ set
underpopulates the region of low color excess values. The differences
in k$^{\prime}$($\lambda$)-k$^{\prime}$(2.2) can therefore be
attributed to the statistical uncertainty associated with the small
number of points on which the linear fits are performed. Hence, we
adopt the dispersion of the two datasets as the actual uncertainty in
our measurements.

For the Galactic diffuse interstellar medium, the total-to-selective
extinction at optical wavelengths is k(0.55)=3.1, and the selective
extinction is k(0.55)$-$k(2.2)=2.8. The mean value of
k$^{\prime}$(0.55)$-$k$^{\prime}$(2.2) is $\sim 2$ for the obscuration
curve of Figure~6, smaller than the Galactic value. The present data
are not suited for deriving the total-to-selective obscuration at
optical wavelengths k$^{\prime}$(0.55), because we are lacking the
reference point at zero obscuration (e.g. radio data). However, since
the total-to-selective obscuration at K, k$^{\prime}$(2.2), is
expected to be small, and probably not larger than the corresponding
Galactic value k(2.2)=0.3, k$^{\prime}$(0.55) should be only a few
tenths larger than k$^{\prime}$(0.55)$-$k$^{\prime}$(2.2). It follows
that the total-to-selective obscuration in the starburst galaxies is
smaller than the total-to-selective Galactic extinction. This method
normalizes the obscuration curve to the color excess of the nebular
gas [equation (4)]; as will be seen in the next section, the shallow
trend of k$^{\prime}$($\lambda$), expecially at UV wavelengths, can be
interpreted as a difference in the covering factors of the dust in
front of the stars and of the gas.

Emission from hot dust may in principle originate around HII regions,
where the high UV energy density may heat a fraction of the available
dust to high temperatures (T$\ge 500$~K, e.g. Panagia 1978), and
contribute to the K-band flux. A contribution to the K-band
proportional to the amount of reddening from the hot dust would bias
the slope $s$ of Equation~2 by increasing its absolute value, and
making it more negative.  However, the K-band flux of the ``dusty''
starburst galaxy M~82 receives less than 15\% contribution from hot
dust (Satyapal et al. 1995). In addition, bluer wavelengths are
progressively less affected by hot dust emission than the K~band, and
our results on k$^{\prime}$($\lambda$) do not change if the J~band
instead of the K~band is chosen as reference point.

Two basic hypotheses are used in deriving k$^{\prime}$($\lambda$): 1)
the stellar continuum ratios F($\lambda_X$)/F(K) are mostly determined
by dust reddening and the variations in the composition of the stellar
population in each galaxy are a secondary effect; 2) the ageing of the
burst population is also a secondary effect.  To test these
hypotheses, the observed continuum ratios F(0.55)/F(0.16) and
F(2.2)/F(0.55) of the 19 galaxies are compared with predictions from
models of stellar populations in Figure~7.  The two extreme cases of a
single, ageing burst population and of a continuous star formation
population, in the age range 1$\times$10$^7$--1.5$\times$10$^{10}$~yr,
are considered (Bruzual \& Charlot 1996, Leitherer \& Heckman 1995,
LH95 hereafter, see also section~6). In neither case, the predicted
colors are able to reproduce the observed ones. The reddening lines
are also shown, using the Galactic and LMC extinction curves and the
curve of Figure~6. The trend of the data points is compatible with
being mostly due to dust obscuration, and the scatter around the
reddening line can be attributed to variations in the intrinsic
populations from galaxy to galaxy. Some scatter is indeed expected due
to the large galaxy region subtended by the observational apertures.
The separate position occupied by NGC~4861 in the diagram is a
consequence of the younger stellar population which characterize this
galaxy relative to the others (see section~7). The location of IC~214
is explained by the inaccurate calibration at optical wavelengths.

Further evidence that the reddening and not the ageing of the burst 
population is the main responsible for the observed trends comes from
the study reported in CKS94. In starburst galaxies, the UV spectral
index is weakly dependent on age; the number of ionizing photons
decreases significantly (and the nebular lines becomes undetectable)
before the intrinsic value of the UV spectral index changes appreciably
 (LH95). Therefore, the correlation of the UV spectral
index with the color excess E(B$-$V) of the gas, found by CKS94, is
an effect of dust reddening; the ultraviolet stellar continuum becomes
redder as the reddening of the gas increases. 

\subsection{A Model for the Reddening}

The distribution of stars, gas, and dust in galaxies is complex enough that in
general it cannot be fully accounted for by simple models. However, models
provide pictorial guidelines on the true distribution, and, in some cases,
can satisfactorily describe the observables, such as the reddening of the 
emerging radiation. 

The foreground clumpy dust model, which is able to describe the
reddening of the nebular gas (CKS96), will be employed to pin down any
difference between the obscuration of the stellar continuum and the
obscuration of the gas, from the UV to the K band.  In the model, the
dust clumps are assumed to be all equal and to be Poissonian
distributed along the line of sight, with average number $\cal N$
(Natta \& Panagia 1984, CKS94). The clumpy distribution is also
assumed to be foreground to the ionized gas, and, therefore, to the
starburst region; however, it is not required that it is foreground to
the entire galaxy.  It should be remembered that the goal of the model
is to provide some insight on the systematic behavior of the reddening
suffered by the stars, but may not necessarily yield the {\em exact}
geometrical distribution of the dust in the galaxy.

The general expression for the color excess of the radiation crossing
a foreground distribution of clumps is given by (CKS94):
$$E(B-V)_{H_a/H_b}= 1.086 {\cal N}^g {\displaystyle {e^{-\tau_c(H_b)}
- e^{-\tau_c(H_a)}}\over \displaystyle k(H_a) - k(H_b)},\eqno(5)$$
where H$_a$ and H$_b$ are defined as in Equation~(1), ${\cal N}^g$ is
the average number of clumps crossed by the nebular radiation,
$\tau_c(\lambda)$ is the optical depth of a clump at wavelength
$\lambda$, which satisfies the relation
$\tau_c(\lambda_a)/\tau_c(\lambda_b)=k(\lambda_a)/k(\lambda_b)$, and
k($\lambda$) is the extinction curve (e.g., Galactic or LMC). Two
cases are shown in Figure~8, overplotted to the data of CKS96, and
correspond to two configurations of the relative source-dust
location. In the first case (dashed line in Figure~8), the dust layer
is close to the radiation source, so light from regions outside the
observational aperture is scattered into the line of sight, and adds a
positive contribution to the observed radiation (scattering model,
this is also the case shown in Figure~1 of CKS96). In the second case
(continuous line), such ``extra'' contribution is not present, because
the clumpy dust layer is distant from the radiation source, or because
the observational aperture is large enough to encompass most of the
scattering regions (non-scattering model). The non-scattering model is
in better agreement with the observed color excess than the scattering
model, and is used to investigate the behavior of the stellar
continuum. The agreement with the data of the non-scattering case is
not surprising, since the observational aperture encompasses a major 
fraction of the galaxy at the typical distance of 60~Mpc.

Since the reddening of the stellar continuum correlates with the
reddening of the nebular gas, the same model can be
employed to describe the obscuration of gas and stars. However, the
possibility that the covering factor of the dust changes from gas to
stars should be included. For a foreground distribution
of clumps, the selective obscuration of the stellar continuum relative
to the emission lines, k$^{\prime}$($\lambda$)$-$k$^{\prime}$(2.2), is then 
given by (CKS94):
$$k'(\lambda)-k'(2.2)=\ f\ {\displaystyle {[e^{-\tau_c(2.2)} - 
e^{-\tau_c(\lambda)}]}\over \displaystyle [e^{-\tau_c(H_b)} - 
e^{-\tau_c(H_a)}]}\ [k(H_a)-k(H_b)],\eqno(6)$$
where $f={\cal N}^s/{\cal N}^g$ is the ratio of the average number 
of clumps in front of the stars to the number in front of the ionized gas 
(i.e., the ratio between covering factors). 
The shape of the curve in Equation~(6) depends on the optical depth of the 
clumps, and on the two emission lines H$_a$ and H$_b$ selected for the 
normalization; however, for our range of optical depths, 
$\tau_c(0.55)< 3.7/{\cal N}^g$ or, if ${\cal N}^g> 10$, 
E(B$-$V)$_{H\beta/Br\gamma}<$1, such dependences are not important. 

The effect of the parameter $f$ on the selective obscuration described
by Equation~(6) is shown in Figure~9. If both the ionized gas and the
stars cross the same dust layer, $f=1$; the case shown in the figure
(curves 1G and 1L) corresponds to ${\cal N}^g$=10 and
E(B$-$V)$_{H\beta/Br\gamma}$=1.0, which implies
$\tau_c(0.55)=0.37$. The predicted optical and UV rise is clearly
steeper than the observed one, and the disagreement is not lessened if
the model parameters are changed: the dependence on $\tau_c$ is weak
for our range of values, and there is no explicit dependence of the
curve on the average amount of clumps $\cal N$. However, changing the
value of $f$ changes the normalization of the obscuration curve. A
qualitative agreement between data and models is observed in Figure~9,
for both the Galactic and the LMC extinction curves, if $f=0.6$ is
adopted (curves 2G, 2L, 3G, and 3L).  From a quantitative point of
view, small disagreements still exist between data and models: the
Galactic curve has in general too strong a 0.22~$\mu$m feature, and
the LMC curve is too steep in the far-UV, relative to the CKS94
curve. These deviations are accentuated for small values of
$\tau_c(0.55)$, although in such regime the reddening effect of the
dust on the emerging radiation is small, and the details of the
selective obscuration are lost in the measurement uncertainties.

Physically, ${\cal N}^s/{\cal N}^g=0.6$ means that the stellar
continuum radiation crosses only 60\% of the clumps crossed by the
nebular radiation before emerging from the galaxy. The covering factor
of the dust in front of the gas is then almost twice as much as the
covering factor of the dust in front of the stars. The dust
distribution is not only clumpy, but is also unevenly distributed
in front of stars and gas, and the ionized gas is, on average, more
closely associated with the dust than the stars are (cf. CKS94; see,
however, Keel 1993). These results on the reddening of the stellar
continuum in starbursts will be applied in sections 6 and 7 to derive
the intrinsic shape of the spectra.

\section{The Iron Emission and the Supernova Rate}

A first constraint on the star formation histories of our galaxies can
be obtained from the comparison of the present supernova rate, which
is a measure of the past star formation activity, with the present
rate of production of ionizing photons, which is a measure of the
current star formation activity (Kennicutt 1983). The infrared [FeII]
emission line can be used to derive the present supernova rate
(Moorwood \& Oliva 1988, Greenhouse et al. 1991). The [FeII] line at
1.257~$\mu$m and the molecular hydrogen emission H$_2$~v=1--0~S(1) at
2.121~$\mu$m were detected in 14 of the galaxies listed in Table~3.

The [FeII] emission is believed to be linked to supernova activity
(Moorwood \& Oliva 1988, Greenhouse et al. 1991, Forbes \& Ward 1993,
Mouri, Kawara, \& Taniguchi 1993, van der Werf et al. 1993, Armus et
al.  1995). Iron is a refractory element heavily depleted on dust in
the diffuse ISM (e.g. De Boer, Jura \& Shull 1987). Fast shocks
propagating in the diffuse ISM (v$\ge$100~km/s), as those produced by
supernova explosions or by superwinds, can destroy dust grains through
sputtering processes or grain-grain collisions, and replenish the ISM
with gas-phase iron (e.g. Draine 1990). The gaseous iron is then
collisionally excited in the cooling post-shock gas, and produces the
observed infrared emission (Shull \& Draine 1987). Supernova remnants
show enhanced [FeII]/H ratios, up to about a factor 1000 more than HII
regions (e.g., Graham, Wright \& Longmore 1987, Oliva, Moorwood \&
Danziger 1989). In galaxies, the [FeII] emission appears to be
positionally coincident with the radio emission from SNRs (Greenhouse
et al. 1991, Forbes et al. 1993), and with regions of star formation
(van der Werf et al. 1993, Armus et al. 1995).

Unlike the case of [FeII], more than one mechanism can produce the
infrared emission from molecular hydrogen, and both nonthermal
processes, like fluorescent excitation of H$_2$ (Black \& van Dishoeck
1987), and thermal processes are involved.  In the thermal processes,
collisionally excited H$_2$ is present on the surface of molecular
clouds heated by shocks, or UV and transient soft X-ray radiation
(Draine, Roberge, \& Dalgarno 1983, Hollenbach \& McKee 1989,
Sternberg \& Dalgarno 1989, Draine \& Woods 1990). Such processes
appear to be the dominant mechanism for producing the H$_2$~v=1-0S(1)
emission in galaxies (Moorwood \& Oliva 1990, Elston \& Maloney 1990,
Israel \& Koornneef 1991, Kawara \& Taniguchi 1993, Forbes \& Ward
1993, van der Werf et al. 1993). In thermal processes, the line ratio
1--0S(0)(2.22~$\mu$m)/1--0S(1)(2.12~$\mu$m) is predicted to be $<$0.4
(Mouri 1994); the upper limits on the line ratios of our galaxies
(Table~4) are consistent with this requirement.  Shocks from supernova
explosions, superwinds, and collisions of interstellar media in
mergers (Rieke et al. 1985, van der Werf 1993) compete with the UV
radiation from OB associations to excite H$_2$ in galaxies (e.g,
Puxley et al. 1990, Israel \& Koorneef 1991). Slow shocks
(v$<$50~km/s) are required to excite H$_2$ (Draine et al. 1983),
because fast shocks dissociate it. However H$_2$ can form again in the
cooling post-shock gas of a fast shock (Hollenbach \& McKee 1989); in
addition, shocks can be slowed down in the transition from the low
density diffuse ISM into the high density molecular cloud (van der
Werf et al. 1993, and references therein).

The distribution of the line intensities of [FeII] and H$_2$ relative
to the hydrogen recombination lines is shown in Figure~10. Most of the
starbursts crowd around the values [FeII]/Pa$\beta\sim$0.3 and
H$_2$/Br$\gamma\sim$0.5, and have [FeII]/H$_2$ ratios in the range
2--6 (cf.  column~6 of Table~4), in agreement with previous results
(Moorwood \& Oliva 1988, Greenhouse et al. 1991, Goldader et
al. 1995). The ratio [FeII](1.257~$\mu$m)/Pa$\beta$ can be related to
the more widely used ratio [FeII](1.644~$\mu$m)/Br$\gamma$ by using
[FeII](1.257~$\mu$m)/[FeII](1.644~$\mu$m)=1.36 from the atomic data of
Nussbaumer \& Storey (1988), and Pa$\beta$/Br$\gamma$=5.72 from case B
recombination with electron temperature T$_e$=7500~K and density
n$_e$=100~cm$^{-3}$ (Osterbrock 1989). The [FeII]/H$_2$ values are
corrected for intrinsic reddening, using the color excess from the
Pa$\beta$/Br$\gamma$ ratio and adopting the Galactic extinction curve
(CKS96), with the assumption that [FeII], H$_2$, and H are affected by
the same obscuration. The LINER NGC~6764 has ratios
[FeII]/Pa$\beta$=1.06 and H$_2$/Br$\gamma$=2.62, larger than the
typical values of starburst galaxies, but compatible with galaxies
with active nuclei and with mergers (see, e.g., the compilation in
Greenhouse et al. 1991, and van der Werf et al. 1993). 
In a flux-flux diagram, the correlation of H$_2$ with the H recombination 
lines is less tight than the correlation of [FeII] with the H lines (see 
Figures~11 and 12). This supports the idea that the excitation mechanisms 
of H$_2$ are not directly related to the excitation mechanisms of hydrogen 
or [FeII], but are likely a mixture of more than one process (cf. Forbes \& 
Ward 1993, van der Werf et al. 1993, Moorwood \& Oliva 1994, Mouri 1994). 

The [FeII] emission from a galaxy may be thought of as a measure of
the ``current'' supernova rate (Greenhouse et al. 1991), since the
[FeII]-emitting lifetime of a single supernova remnant is believed to
be rather short, $\sim$10$^4$~yr (Oliva et al. 1989). The nebular H
emission can be used to derive a ``predicted'' Type II supernova rate,
because it gives a measure of the number of massive stars which are
presently ionizing the gas. The total number of Type II supernovae
(SNII) produced by an instantaneous burst of star formation is at most
a factor 2 lower than the total number of Type Ia supernovae (SNIa),
but is spread over timescale about 30 times shorter (Greggio 1996). It
follows that in a steady star formation regime, the SNII rate is about
one order of magnitude larger than the SNIa rate, as observed for late
type spirals (van den Bergh \& Tamman 1991). In this case, the [FeII]
emission is produced mainly by shocks from SNII (and from SNIb, if
these arise from massive stars, see Ratnatunga \& van den Bergh 1989),
and is a measure of star formation activity in the recent past.

The tight correlation between [FeII] and the H recombination lines for
the starburst galaxies in this sample (with the exception of NGC~4861
and NGC~1569, see below) suggests that the current supernova rate
yields numbers which are proportional to the predicted supernova
rate. The galaxies are characterized by similar values of the oxygen
abundance (CKS94), therefore metallicity effects should not be
important here. In an instantaneous burst of star formation, the
number of ionizing photons decreases by 2 orders of magnitude after
7--8~Myr, while the first Type II supernova appears after 3--5~Myr
(LH95). It is unlikely that all the galaxies in this sample are in the
narrow evolutionary window between 4 and 8 Myr, and we favor
continuous star formation to explain the tight correlation of
Figure~12. This merely says that the spatial region sampled by our
spectra is large enough, being typically $\sim$4.5~kpc in diameter, to
encompass more than one star-forming knot (cf. Meurer et al.  1995),
and the star formation histories of individual knots are averaged into
a ``global'' star formation event.

Continuous star formation is also supported by the small dispersion
around the mean shown by the reddening-corrected UV/K stellar
continuum ratio (see Table~5 and Figure~5a). The largest observed
dispersion is about $\pm$0.2~dex, while the UV/K ratio of a stellar
population generated by an instantaneous burst of star formation
changes by 1.7~dex over 10$^7$~yr, for a Salpeter IMF with upper mass
limit 100~M$_{\odot}$ (using the LH95 models with solar metallicity).
The variation in the UV/K ratio is smaller, about 0.7~dex, for a
timescale of 5$\times$10$^6$~yr, that is, before the supergiant stars
begin to contribute to the K-band flux. If the starburst events in our
galaxies were generated by instantaneous bursts, they should have
fine-tuned ages, in order to be all observed within the narrow range
of a few Myr.

The correlation between [FeII] and Pa$\beta$ is fit by:
$$[FeII](1.257~\mu) = 0.34\ Pa\beta.\eqno(7)$$
Assuming a mean [FeII] luminosity of a few times 10$^3$~L$_{\odot, bol}$ for 
a supernova remnant (van der Werf et al. 1993, Oliva et al. 1989), 
the current supernova rate of the galaxy can be written:
$$snr([FeII]) = {\displaystyle{L_{[FeII]}/L_{\odot, bol}}\over
\displaystyle 6.6 \times 10^7 (E_{\circ}/10^{51} ergs)},\eqno(8)$$
where the formula used by van der Werf et al. (1993) has been adapted
to the iron emission at 1.257~$\mu$m, with the [FeII] luminosity
expressed in units of the bolometric solar luminosity, and E$_{\circ}$
is the mean energy released in the supernova explosion (see, also,
Colina 1993). The hydrogen recombination lines can be used to derive a
star formation rate (SFR) and to predict a supernova rate. From LH95,
with a Salpeter IMF, the SFR derived from the Br$\gamma$ line is:
$$SFR100 = 6.17 \times 10^{-40}\ L(Br\gamma),\eqno(9a)$$
for a stellar mass range 0.1--100~M$_{\odot}$, and:
$$SFR30 = 2.67 \times 10^{-39}\ L(Br\gamma),\eqno(9b)$$
for a stellar mass range 0.1--30~M$_{\odot}$.
The predicted supernova rate for a minimum mass of 8~M$_{\odot}$ for a star to 
explode is (Elson, Fall \& Freeman 1989): 
$$snr\simeq 0.006\ \Gamma\ SFR.\eqno(10)$$ 
$\Gamma$ is the ratio of the SNII$+$SNIa rate to the SNII rate (and
possibly SNII$+$SNIb rate, if the SNIb have massive progenitors); this
ratio is close to 1 for our purposes.  Table~6 lists for each galaxy
the [FeII]~(1.257~$\mu$m) luminosity, the SFR100, the SFR30, and the
three supernova rates derived from the iron luminosity, and from the
two star formation rates, respectively. The predicted supernova rates
from the SFRs bracket the current supernova rate derived from the
[FeII] luminosity, consistent with the continuous star formation
scenario. The agreement between the various snr estimators is
reasonable if uncertainties in the energy released by supernovae, in
the intrinsic [FeII] luminosity of each supernova, in the SNII/SNIa 
ratio, and in the IMF used to derive the SFR are accounted for.

The [FeII]/H ratios of NGC1569 and NGC4861 are one order of magnitude
or more below the average (see Figures~10 and 12). The H$_2$/H ratios
are also low, although, given the large spread in the values of the
other galaxies, they are not as remarkable as the iron (see
Figure~11).  NGC~4861 and NGC~1569 are the two nearest galaxies in the
sample, and are two metal-poor dwarfs, with oxygen abundance
$12+\log(O/H)\simeq8.3$ (Storchi-Bergmann et al. 1994) or lower (Barth
et al. 1994). Metallicity effects can explain the low [FeII]/H ratios
only if the [Fe/O] abundance ratio is altered (see Gilmore \& Wyse
1988), since the two galaxies have on average 1/3 of the oxygen
abundance of the other galaxies, and more than one order of magnitude
difference in the average [FeII]/H.  The NIR recombination lines of
NGC~4861 are dominated by the emission from the Bright Knot, which is
a high ionization region with an estimated age of $\sim$4.5~Myr (Barth
et al. 1994) and electron temperature and density T$_e\sim$14,000~K
and n$_e\sim$100~cm$^{-3}$ (Dinerstein \& Shields 1986). The
photoionization equilibrium thus favors ionization states of iron
higher than [FeII] (Greenhouse et al. 1991), and may also account for
the lower than average H$_2$ emission (Puxley et al. 1990). In
addition, the first supernova explodes around 3-5~Myr, depending on
the IMF upper mass limit (Leitherer, Robert \& Drissen 1992). Few
supernovae may have been produced by the current star formation
episode, and the mechanical energy released into the ISM may be still
insufficient to free a significant amount of iron into the
gas-phase. Other HII regions which surround the Bright Knot are not as
bright (Barth et al. 1994), and do not contribute significantly to the
integrated properties. At the distance of NGC~1569 (about 2.5~Mpc,
O'Connell, Gallagher, \& Hunter 1994), the NIR spectroscopic aperture
subtends a region of 240$\times$85~pc$^2$. A close inspection of the
NIR spectra of NGC~1569 reveals contributions from two knots, which
are identified with the postburst super-star cluster A (Hodge 1974,
O'Connell et al. 1994) and with the bright HII region ``2'' (Waller
1991). The HII region ``2'' is responsible for $\simeq$85\% of the
line emission in the NIR spectra, while Region~A provides $\simeq$65\%
of the continuum. The HII region is presently among the most active
sites of star formation in the galaxy, with an upper limit on the age
$\sim$5~Myr (Waller 1991). Region~A has been the site of a powerful
star formation episode about 15~Myr ago (O'Connell et al. 1994), and,
together with Region B, is now driving a gas outflow (Heckman et
al. 1995b). For the HII region, considerations similar to NGC~4861
apply. For Region~A, the superwind generated by the starburst may have
already ejected from the small region subtended by our aperture
significant amounts of interstellar gas and dust (Marlowe et al. 1995,
Heckman et al. 1995b).

\section{Star Formation History}

\subsection{The Template Starburst Spectrum}

The reddening correlations of section~4.1 provide a method to correct the 
spectral energy distribution of the starburst galaxies for the effects of 
dust obscuration. The reddening-corrected flux densities can then be used 
to study the star formation histories of the galaxies in the sample. 

Because of the large observational aperture employed, the observed
spectra must receive a sizeable contribution from stellar populations
other than the one produced by the current star formation event.
Indeed, the values of the EWs of the hydrogen emission lines are
smaller than expected from a single young stellar population. The
average values are 100~\AA, 18~\AA, and 14~\AA, for H$\alpha$,
Pa$\beta$, and Br$\gamma$, respectively, with dispersions around the
means of about 60\%. The H$\alpha$ EWs are corrected for the
differential reddening between lines and continuum, using the
multiplicative factor $10^{0.4 E(B-V) [k(\lambda)-k'(\lambda)]}$; the
IR lines are only mildly affected by this problem (see Table~5). If
the star formation episode has been sustained at a constant rate over
a few million years, as suggested by the comparison between the
current and the predicted supernova rate in the previous section, 
 then the expected values for the
EWs of the hydrogen emission lines are much larger than measured, with
EW(H$\alpha$)$>$500~\AA, EW(Pa$\beta$)$>$100~\AA, and
EW(Br$\gamma$)$>$60~\AA (from LH95, adapted to a Salpeter IMF in the
mass range 0.1--100~M$_{\odot}$).

The comparison of the observed galaxy spectral energy distributions
with models provides further constraints on the stellar populations
contributing to the emerging radiation. Since we are interested in the
{\it mean} stellar population(s) responsible for the relatively tight
correlations of Figure~5, the zero-reddening flux density ratios
reported in Table~5 can be used to create a {\it template} starburst
spectrum (TSS) in the UV-to-NIR wavelength range. To improve the
resolution in the wavelength range 0.125-1.0~$\mu$m, the mean spectra
of CKS94, rather than the broad band data of Table~5, are used for the
TSS, after correction for reddening using the curve in Figure~6, and
after the spectra are averaged to give a single zero-reddening
template. The UV-optical template is therefore produced from a sample
of 33 starbursts (CKS94), with a spectral resolution of 10~\AA, and is
normalized relatively to the {\it mean} K-band flux density using the
F(0.70)/F(2.2) ratio as reference point. A direct comparison with the
broad band data shows that the two sets of data are consistent with
each other, and describe the same mean spectrum. Because of their
resolution, the spectra from CKS94 provide detailed information on the
far UV rise and the amplitude of the $\sim$0.4~$\mu$m discontinuity,
which may help constrain the stellar populations responsible for the
observed radiation. The TSS is shown in Figure~13.

The spectral synthesis models of Bruzual \& Charlot (1996, BC96
hereafter) are used to generate spectra for the fit of the TSS. The
adopted IMF is Salpeter in the mass range 0.1--100~M$_{\odot}$. The
BC96 models do not include the contribution from the nebular continuum
emission, which we obtain from LH95 and add to the spectra produced by
the BC96 code at the appropriate age. The BC96 models are given for
solar metallicity only; this is sufficiently adequate for the present
sample, where most of the galaxies have about solar metallicity. The
synthetic spectrum to be compared with the TSS is built by combining
three stellar populations of increasing age. The weights of the three
populations are left as free parameters which are fitted via the
minimum $\chi^2$ technique. The best fit model must also satisfy the
additional constrain of being able to reproduce the observed hydrogen
line EWs. Of the three stellar populations, one is assumed to be young
and responsible for most of the ultraviolet continuum (``star-forming
population''); the other two are assumed to be older than the first
and contribute mainly to the optical and IR continuum, to dilute the
hydrogen line EWs (``underlying populations''). For this reason, the
star-forming population is selected in the age range
0--3$\times$10$^7$~yr, and both the cases of an instantaneous burst of
star formation and of a constant SFR regime are considered. The
underlying populations have ages $>$7$\times$10$^7$~yr, and are
produced by bursts of star formation. The latter is not a restrictive
assumption if the the burst duration is finite, but much shorter than
the age of the stars. A second type of fit of the TSS is also
attempted by using only one star-forming population with constant SFR,
and leaving the age as free parameter.

The spectrum produced by a star-forming population alone has
systematically bluer colors than the TSS at UV wavelengths, for a
stellar age of less than a few 10$^7$~yr. Figure~13a shows the example
of a stellar population with constant SFR and 2$\times$10$^7$~yr
old. This spectrum is clearly bluer than the TSS at all
wavelengths. Among the instantaneous-burst populations, the one which
comes closer to the TSS has an age of 10~Myr, still a much bluer UV
spectrum, and far too red optical$-$IR colors. The TSS therefore
appears to receive additional contribution from stellar populations
which are older than a few times 10$^7$~yr. A starburst galaxy may
indeed be conceived as a high--SFR event superimposed to an underlying
quiescent galaxy, which is probably forming stars at a roughly
constant (and small) rate since a Hubble time. In Figure~13b, this
galaxy is modelled by the combination of two stellar populations: the
first is undergoing star formation at a constant rate since
1.5$\times$10$^{10}$~yr and represents the underlying galaxy
population; the second is undergoing star formation at a constant rate
since 2$\times$10$^7$~yr and represents the starburst population. The
fit of this model to the TSS is quite unsatisfactory, expecially at
optical wavelengths, where the model spectrum tends to be redder than
the TSS. This is a common characteristics to all the synthetic spectra
obtained by combining a relatively young population (age
$\sim$10$^7$~yr or younger) with an old underlying population (age
$\sim$10$^{10}$~yr).

A satisfactory fit of the TSS can be obtained with a stellar
population characterized by constant SFR since 1--2$\times$10$^9$~yr
(see Figure~13c). The low mass stars produced during this period of
time, and which contribute mostly to the optical and IR flux, account
approximately for the dilution of the EWs of the hydrogen emission
lines, although a detailed comparison would imply an heavy
extrapolation from the LH95 diagrams. The HI gas reservoir of our
galaxies may be sufficient to support such long star formation
timescales at the present SFR (columns 2 and 4 of Table~8).  For
Mrk~357, NGC~1614, NGC~4194, NGC~4385, NGC~6090, and Mrk~309, whose
gas depletion timescale is shorter than 10$^9$~yr, the present burst
of star formation is probably an exceptional event. However, molecular
hydrogen has not been included in the gas supply estimates of Table~8,
and, if present in large amounts, would considerably increase the
depletion times. Rather than a single large star formation event
lasting 10$^9$~yr, the starburst region subtended by our large aperture
 may be thought of as a collection of more
modest bursting knots characterized by different values of the SFR and
igniting at different epochs of the galaxy's life. A way to reduce the
SFR and allow the burst to last longer with the available gas supply
is to assume that low~mass stars are not produced during bursts of
star formation, or, alternatively, that the IMF is flatter than
Salpeter below a certain stellar mass; if the low mass limit for the IMF is
1~M$_{\odot}$, instead of 0.1~M$_{\odot}$, the SFRs decrease by a
factor $\sim$2.5, for $\alpha$=2.35. Whether this is the case in
starburst galaxies is still debated (e.g., Rieke et al. 1980, 1993,
Lester et al. 1990, Satyapal et al. 1995, BernL\"or 1993; see also
below).

A better fit to the data in terms of $\chi^2$ is given by the
combination of three components: a 2$\times$10$^7$~yr population with
constant SFR, and two intermediate age populations, generated by
instantaneous bursts 10$^8$~yr and 5$\times$10$^8$~yr ago,
respectively. They contribute 15\%, 13\%, and 72\%, respectively, of
the K-band flux (see Figure~13d). The intermediate age populations are
responsible for the dilution of the EWs of the hydrogen emission
lines,  and account for the small values of the observed H$\alpha$,
Pa$\beta$, and Br$\gamma$ EWs (see Table~7). An additional consistency
check is to compare the K-band luminosities of the galaxies with the
model. Column~7 of Table~8 gives for each galaxy the reddening
corrected K-band luminosity, L$_K$, in units of the solar K-band
luminosity, and normalized to the star formation rate SFR100; these
values are to be compared with the quantity
5.28$\times$10$^9$~L$_{\odot}$~yr/M$_{\odot}$ derived for the
model. Most of the galaxies lay within a factor two of the model
value, and the median value for the observed L$_K$/SFR100 is
5.94$\times$10$^9$~L$_{\odot}$~yr/M$_{\odot}$.  The exceptions are
NGC~4861 and NGC~6217. As will be seen in a following section, the
K-band flux of NGC~4861 receives about 50\% contribution from the
bursting population; in the case of NGC~6217, it is likely that old
stellar populations provide the bulk of the K-band flux. 
The mass in stars per unit of SFR predicted by the model is
1.62$\times$10$^9$~M$_{\odot}$. This value is also the timescale over
which the stars have been produced, namely 1.62$\times$10$^9$~yr, if
the SFR remained constant over that period of time; this is in rough 
agreement with the age of the oldest stellar population from the model
($\approx$500~Myr). If the star formation is episodic, some
of the past episodes may have been characterized by different
(possibly higher) values of the SFR, and the entire star-forming event by 
a different (possibly shorter) timescale. 
The amplitude of the discontinuity at 0.4~$\mu$m in the synthetic
model is overpredicted relative to the observations; this
characteristic remains true also when model fitting of individual
galaxies is attempted (see below), and may be partially explained by
the fact that the wavelength region 0.32--0.38~$\mu$m is the noisiest
in our spectra (cf. McQuade et al. 1995, Storchi-Bergmann et
al. 1995). As already discussed above, underlying populations older
than 1--2~Gyr produce increasingly worse fits of the TSS, mainly because
the synthetic spectra are far too red at optical wavelengths; the
underlying stellar population contributing to the mean starburst
spectrum must thus be younger than $\approx$2~Gyr.  Changing the
metallicity of the model populations from solar to half solar (the
typical value of our galaxies) has a negligible effect on this
conclusion.

The small values of the hydrogen line EWs can be explained either as a
dilution effect from an underlying population or as evidence for a
high-mass-star~truncated IMF.  A 30~M$_{\odot}$ upper mass limit
produces a factor 4--5 less ionizing photons than a 100~M$_{\odot}$
upper mass limit, for a Salpeter IMF (LH95), and therefore, smaller
emission line EWs. For instance, a $\sim$5$\times$10$^7$~yr
star-forming population, with constant SFR and a 0.1--30~M$_{\odot}$
Salpeter IMF, accounts for the observed mean EWs of Table~7. However,
its spectrum does not reproduce other observables; it is, for
instance, much bluer than the TSS (Figure~14). A fit of both the TSS
and the line EWs with a high-mass~truncated IMF still requires the
presence of intermediate age underlying populations. In this case, the
best fitting star-forming population has age 0.5--1$\times$10$^7$~yr,
which is when red supergiants appear; a small change in age thus
corresponds to a variation up to a factor 20 in the EW of the infrared
hydrogen lines (LH95). This is in contradiction with the observational
values of the Br$\gamma$ EW, which show a small spread around the mean
value.  The model predicts a K-band luminosity per unit of SFR30
$\sim$3.0$\times$10$^9$~L$_{\odot}$~yr/M$_{\odot}$, which is a factor
$\sim$2 larger than the observed mean value
L$_K$/SFR30$\simeq$1.4$\times$10$^9$~L$_{\odot}$~yr/M$_{\odot}$. This
prediction is minimally dependent on the age of the star-forming
population, since the intermediate age populations contribute more
than 80\% of the K-band flux, even when the red supergiants appear.
The gas consumption timescales decrease by a factor 4.35, going from
M$_{up}$=100~M$_{\odot}$ (cf. Table~8) to M$_{up}$=30~M$_{\odot}$; for
most of the galaxies in the sample the HI gas depletion time is, then,
less than the age of the older population predicted by the
model. There is, in conclusion, no compelling evidence for a
high-mass-star~truncated IMF from this set of data (see, also, Moffat
1996).

In summary, the TSS can be described either by a single star-forming
population which is undergoing constant star formation since
1--2$\times$10$^9$~yr, or by the combination of a relatively young
($\sim$2$\times$10$^7$~yr), constant SFR population with two older
underlying populations, in the age range 10$^8$--10$^9$~yr and 
no longer forming stars. The fitting technique is not sophisticated
enough to break the two intermediate age populations into more
components. It is interesting, however, that at least two intermediate
age components, and maybe a continuously star-forming population, are
needed to fit the observed spectrum; this support the idea that
starburst activity in galaxies is spread over a timescale of roughly
10$^9$~yr and proceeds through episodes which may involve different
but spatially close regions (e.g. Waller 1991).  Stellar populations
older than $\sim$10$^9$~yr are not needed to account for the TSS,
implying that the emission from the most recently formed stars
dominates the observed spectrum not only in the UV, but also at
optical and IR wavelengths. This explains, a posteriori, why the 
continuum emission from the galaxies follows a common reddening trend 
at all wavelengths, as seen in section~4.1. 

\subsection{The Mass-to-Light Ratio}

The best fit model to the TSS yields a stellar mass-to-light ratio,
M/L$_K$, of 0.31~M$_{\odot}$/L$_{K, \odot}$, where L$_K$ is the K-band
luminosity; 98.7\% of the mass in this ratio is provided by the two
intermediate age populations. The mean value of the observed
mass-to-light ratio is 1.44~M$_{\odot}$/L$_{K, \odot}$ (see the last
column of Table~8), or about 5 times the stellar mass-to-light ratio
predicted from the model. Each galaxy's mass-to-light ratio has been
derived from the dynamical mass, M$_d$, subtended by the observational
aperture, using the HI velocity width from Huchtmeier \& Richter
(1989). The difference between the observed and the model
mass-to-light ratios allows room for the presence of molecular and
atomic gas, of stellar remnants and of low~mass~stars (older than
$\sim$2$\times$10$^9$~yr) which contribute negligibly to the observed
luminosity, but will contribute to the mass. An estimate of the HI
mass contained within the observational aperture is given in column~6
of Table~8, and is in general a small fraction of the total mass
within the aperture; the values of column~6 are derived assuming a
spherical distribution for the HI gas. To estimate the contribution to
the dynamical mass of the stars older than $\sim$2$\times$10$^9$~yr,
we assume that the stellar populations follow a Hubble luminosity
profile (Binney \& Tremaine 1987). For a galaxy with typical stellar
mass 10$^{11}$~M$_{\odot}$ and radius 10~kpc, the mass contained
within a region of 4~kpc in diameter is about
3.6$\times$10$^{10}$~M$_{\odot}$. This value is a factor 4 smaller if
the galaxy follows an exponential profile. If the age of the stars is
10~Gyr or larger, the K-band luminosity produced by this low-mass~star
population is less than 15\% of the observed mean value, and of the
same order of magnitude of the observational uncertainties.  The total
mass produced by a star formation event lasting 2$\times$10$^9$~yr and
with mean SFR100=9~M$_{\odot}$/yr (as derived from Table~6) is
1.8$\times$10$^{10}$~M$_{\odot}$, or about 1/2 of the pre-existing
stars. Accounting for the mass returned to the ISM by supernova
explosions would introduce a small difference in this estimate,
because most of the mass is locked in low~mass stars.  The total stellar 
mass-to-light ratio is therefore 3 times larger than the mass-to-light
ratio of the population generated by the star formation event, 
 namely $\sim$0.9~M$_{\odot}$/L$_{K, \odot}$.  Only NGC~1614,
NGC~4194, and NGC~4385 have mass-to-light ratios significantly smaller
than 0.9~M$_{\odot}$/L$_{K, \odot}$.  A few reasons may explain their
case: 1) the TSS does not match perfectly each individual spectral
energy distribution, and specific stellar population models should be
applied to individual galaxies; 2) the dynamical mass may be
underestimated, because detailed density profiles have not been taken
into account.

In summary, the mean mass-to-light ratio derived from the observations is
compatible with the value derived from the model. The implication is that an
IMF with lower and upper mass limits in the range 0.1--100~M$_{\odot}$ can
satisfactorily model the data, and a low-mass-stars~deficient IMF is not
required to explain the observations. 

\section{Individual Galaxies: NGC~7714 and NGC~4861}

In order to test the effectiveness of our crude fitting technique, we
try to reproduce the observed spectra of two galaxies: the starburst
prototype NGC~7714 (Weedman et al. 1981), and the bluest galaxy in our
sample, NGC~4861.

The spectrum of NGC~7714 (see Figure~15) is corrected for reddening
using E(B$-$V)=0.4 and the reddening values of Table~5 and CKS94. The
best fit for the spectrum and the emission line EWs is given by a
1$\times$10$^7$~yr population with constant SFR and two underlying
populations with ages 1$\times$10$^8$~yr and 5$\times$10$^8$~yr. The
flux from the three stellar populations accounts for 14\%, 17\% and
69\%, respectively, of the K-band emission. The observed and predicted
emission line EWs are shown in Table~7. The results from this model
suggest that NGC~7714 has been experiencing episodic or continuous
star formation over the last 5$\times$10$^8$--1$\times$10$^9$~yr, as
already concluded by Gonzalez-Delgado et al. (1995). If the SFR were
to remain constant over a period of 10$^9$~yr, the total mass produced
by the star formation event would be
$\sim$6$\times$10$^9$~M$_{\odot}$, comparable to the total HI mass in
the galaxy. The K-band luminosity per unit of star formation rate and
the stellar mass-to-light ratio predicted by the model are
2.67$\times$10$^9$~L$_{\odot}$~yr/M$_{\odot}$ and
0.32~M$_{\odot}$/L$_{\odot}$, respectively, to be compared with the
values shown in Table~8. The observed M/L is about a factor 3.5 larger
than the model M/L, implying that the use of an IMF with mass range
0.1--100~M$_{\odot}$ is appropriate also for the case of NGC~7714. A 
low-mass-stars~deficient IMF does not seem required for this galaxy, unlike 
previous results (Bernl\"ohr 1993, and references therein). Much of
the difference between the present and previous conclusions on the IMF
can be attributed to the reddening corrections adopted by different
authors for the luminosity. If the dust and stars are assumed to be
mixed (cf. Bernl\"ohr 1993), the reddening corrected luminosities will
be larger, and the M/L values smaller, than those obtained by assuming
that the dust is foreground to the stars (cf. Satyapal et al. 1995,
for the case of M82). For instance, in a dust-star mixed geometry a
measured E(B$-$V)=0.22 at optical wavelengths corresponds to a K-band
optical depth $\tau_K=2$ and the IR stellar continuum must be
corrected for a reddening factor 2.3. In a foreground geometry the
same value of the optical color excess corresponds to a K-band optical
depth $\tau_K=0.06$ and to a reddening correction factor of only 1.06
for the IR stellar continuum.

NGC~4861 strongly deviates from the average behavior of the other
galaxies: its spectrum is the bluest in our sample, and the EWs of its
hydrogen emission lines are about one order of magnitude larger than
the others (cf. Tables~2 and 3). NGC~4861 is the closest approximation
to a pure starburst population in the present sample. After reddening
correction, using the same recipe as for NGC~7714 and E(B$-$V)=0.2,
the NGC~4861 spectrum is best fitted by the combination of a
4$\times$10$^6$~yr constant SFR population with a
1.2$\times$10$^{10}$~yr burst population (Figure~16a). The latter
provides 52\% of the K-band flux, 43\% of the J-band flux, and 16\% of
the 0.65~$\mu$m flux.  However, this synthetic spectrum overpredicts
the hydrogen line EWs, expecially in the J and K bands (see model~1 in
Table~7). If we impose that the star-forming population can only
originate from an instantaneous burst of star formation, then the best
fitting synthetic spectrum is given by a combination of a
3$\times$10$^6$~yr and a 1.2$\times$10$^{10}$~yr burst population
(Figure~16b), where the latter provides 62\% of the K-band flux, 52\%
of the J-band flux, and 18\% of the 0.65~$\mu$m flux. The $\chi^2$
from the second model is about 30\% worse than from the first model,
but the predicted values for the hydrogen line EWs are closer to the
observed ones (see model~2 in Table~7). The EWs mismatch between
model~1 and the observations could be alternatively explained if some
of the ionized gas has been blown away from the observed galaxy region
by stellar winds, so it does not contribute to the spectrum. Both
model~1 and 2 are consistent with all the UV flux and most of the
optical flux given by a very young stellar population, in the age
range 3--4$\times$10$^6$~yr, in agreement with previous results (Barth
et al. 1994). In both models, the optical continuum is overpredicted
by $\sim$25--30\%; the observed optical spectrum may actually be
underestimated, because the wavelength region 0.32--0.45~$\mu$m is
missing from the spectrum and a normalization cross-check between the
UV and the optical spectra could not be performed. The use of solar
metallicity models to reproduce the spectrum of this
$\sim$0.1~Z$_{\odot}$ galaxy is not particularly relevant at this
stage of the evolution and for the observables we are interested in,
although it would matter at later stages, expecially around 10$^7$~yr,
when red supergiants appear (LH95).

The UV spectrum of NGC~4861 is slightly bluer than expected from
models (cf.  Figure~16); its spectral index is $-$2.85, to be compared
with a theoretical index of $-$2.7 for a young stellar population
(LH95). Possibly, the galaxy UV spectrum is slightly overcorrected for
reddening. NGC~4861, indeed, does not follow the mean reddening trend
of the other galaxies in the sample (cf.  Figure~5). For this reason,
its spectrum was not used in the derivation of the obscuration curve
or of the Template Starburst Spectrum. The reddening problem is only
worsened by the use of the Galactic or LMC extinction curves, because
these are steeper in the UV than the obscuration curve of Figure~6. In
addition, the Galactic curve overcorrects the 0.22~$\mu$m region (see
Figure~16c). A more complex dust distribution than the foreground
clumpy one adopted here may be present in this galaxy, although the
reddening of the ionized gas does not provide evidence for such case
(CKS96).

\section{Conclusions}

The analysis of the UV-to-NIR spectral energy distribution of 19
galaxies with starburst central regions shows that the reddening of
the stars and of the ionized gas can be described by a foreground
clumpy dust distribution, characterized by different covering factors
in front of stars and gas. In the IR, the reddening of the stellar
continuum is similar in value to that of the ionized gas, but the two
values progressively diverge towards the UV. If a Galactic-type
extinction curve is adopted for the gas, the UV-bright stars are about
60\% less reddened than the nebular gas (cf. CKS94). In terms of a 
clumpy dust distribution, the radiation from the stellar
continuum crosses, on average, 40\% less dust clumps than the
radiation from the gas.  This difference can be understood if the
dust, gas, and stars are not spatially coincident. The large
observational aperture employed for the present sample covers the
central 3--5~kpc in the galaxies, encompassing many regions at
different evolutionary stages. Stellar winds and supernova explosions
can create ``holes'' in the galaxy ISM through which UV--bright stars
can shine. ``Leaky'' star--forming regions may be the source of
ionization of the diffuse gas (Ferguson et al. 1996).  OB stars have
been found in field regions of galaxies, far enough from any bright
HII region that they are likely to be born in situ (Wilson 1990,
Massey et al. 1995). Young stars can be found in regions of very
low density medium, unrelated to the brightest sources of nebular
emission, and physically ``detached'' from the gas they ionize.

A departure from a foreground--equivalent dust distribution at
wavelengths longer than 2.2~$\mu$m is not excluded by the present
results (cf. Puxley \& Brand 1994), since opaque clouds become
optically thin to the radiation of deeply embedded stars at
sufficiently long wavelengths. IRAS data suggest, indeed, that a small
fraction of the ionizing stars may be buried in optically thick clouds
(Calzetti et al. 1995).

The reddening-corrected spectra and broad band flux densities of the
sample galaxies have been combined to form a ``template starburst
spectrum''. This spectrum, which is entirely derived from
observational data, shows the mean intrinsic appearance, in wavelength
space, of a starburst galaxy with 0.5--1 solar metallicity. The fit of
the ``template starburst spectrum'' with stellar population models
requires at least three different components: a star-forming
population, which is responsible for most of the UV emission, and two
intermediate age populations, 1$\times$10$^8$~yr and
5$\times$10$^8$~yr old, respectively, which contribute to a large
fraction of the optical and IR emission. The star-forming component is
producing stars at a constant rate since $\sim$2$\times$10$^7$~yr, in
agreement with the age limits placed to this component by the current
supernova rates, measured from the [FeII] emission line.  The three
stellar components which fit the ``template starburst spectrum'' are
all relatively close in age and younger than 10$^9$~yr. Our fitting
technique is not sophisticated enough to reveal whether more than
three stellar components contribute to the mean spectrum, but the
limiting case of continuous star formation over
$\sim$2$\times$10$^9$~yr produces a synthetic spectrum which is still
in general agreement with the observations. This suggests the that
starburst activity in galaxies is continuous but episodic (cf. Waller
1991) and covers relatively long periods of time, up to
$\sim$1--2~Gyr. Past star formation episodes have been as efficient as
or more efficient than the present one at producing stars, and the
star formation history of a starburst galaxy is characterized by
events with different levels of star formation strength. The galaxy
very old population (age $\sim$10$^{10}$~yr) does not contribute
significantly to the template starburst spectrum, since the best fit
models do not require components older than $\sim$10$^9$~yr. This is
an ``a posteriori'' justification for the use of a foreground dust
distribution to de-redden the galaxy spectra: stellar populations
younger than $\sim$10$^9$~yr may have been generated during star
formation episodes which are associated with the current event.

A Salpeter IMF with mass range 0.1--100~M$_{\odot}$ has been adopted
through the analysis. This assumption has been checked against various
observables, and no obvious contradiction has been found. The observed
galaxy mass-to-light ratios are compatible with a low mass limit of
about 0.1~$M_{\odot}$, and there is no clear need for a
low-mass-star~deficient IMF.  A high-mass-star~truncated IMF is also
in disagreement with the data; a Salpeter IMF in the range
0.1--30~M$_{\odot}$ can account for the small values of the nebular
hydrogen line EWs, but produces a synthetic spectrum which is much
bluer than the ``template starburst spectrum''.

Spectral energy distributions of local galaxies are often compared
with spectra or multi-band photometry of intermediate and high
redshift galaxies in an effort to pin down the stellar composition,
and therefore, the star formation history of the progenitors of
present-day galaxies (e.g., Steidel et al. 1996, Madau et al. 1996).
In this respect, the ``template starburst spectrum'' provides a
baseline for comparisons with high-redshift star-forming galaxies,
since it represents the the reddening-corrected UV-to-IR spectral
energy distribution of a ``typical'' starburst galaxy, as derived from
observations. However, metallicity differences may not be negligible
between present-day and high-redshift galaxies, and comparisons should
take this caveat into account.

\acknowledgments

The author would like to thank Tim Heckman for many discussions and
useful suggestions while this work was taking shape; Anne Kinney for
comments and constant encouragement; Claus Leitherer for providing the
LH95 starburst models; Ron Allen, Ken Freeman, Nino Panagia, Jim Pringle, 
and Rosemary Wyse for a critical reading of the manuscript; Lee
Armus for guidance through the infrared world. The author also
acknowledges hospitality from the Observatories of the Carnegie
Institute of Washington during part of the work. This research has
made use of the NASA/IPAC Extragalactic Database (NED) which is
operated by the Jet Propulsion Laboratory, California Institute of
Technology, under contract with the National Aeronautics and Space
Administration.  Part of this research was supported by the NASA Grant
NAGW-3757.

{}

\clearpage

Fig. 1 -- Infrared spectra at J (left panels) and K (right panels) of 10
galaxies in the sample. The ordinate axis is the flux density F($\lambda$) in 
units of 10$^{15}$~erg/s/cm$^2$/\AA. The position of the emission lines
[FeII]~(1.257~$\mu$m), Pa$\beta$~(1.282~$\mu$m),
H$_2$~$v=1-0S(1)$~(2.121~$\mu$m), and Br$\gamma$~(2.166~$\mu$m) is indicated
for the nearby galaxy NGC~1569 and for the distant galaxy NGC~1614. None of the
emission lines is resolved. The spectra have been smoothed by a 3 pixel boxcar.
\\
\\
Fig. 2 -- Two examples of ultraviolet, optical, and near-infrared spectra with 
no normalization between the three wavelength ranges. The galaxies are 
NGC~1614 and NGC~7714. The flux density is in erg/s/cm$^2$/\AA\ and is plotted 
as a function of the wavelength $\lambda$ in the range 1,200--22,500 \AA.
\\ 
\\
Fig. 3 -- The spectral index $\beta_{im}$ derived from the J, H, and K images 
versus the index $\beta_{sp}$ derived from the NIR spectra. The 1~$\sigma$ 
error bars are shown for each point. As expected, the data follow a straight 
line with slope unity. IC~1586 and Mrk~542 are exceptions to this trend (see 
text). 
\\
\\
Fig. 4 -- The color-color diagram for the 24 galaxies in the
sample. The values are corrected for the foregrond Galactic
extinction. The open squares are the 19 starburst galaxies and the
filled squares are the Seyfert~2s NGC262, NGC1275, NGC1667 and the 
LINER NGC6764, with photometry obtained in a 10$^{\prime\prime}$ circular
aperture (Table~1); the position of NGC4861 is labelled. The filled
triangles mark the trajectory of the colors of NGC~1068 for increasing
extraction apertures (in the direction indicated by the arrow). The
apertures are 5$^{\prime\prime}$, 10$^{\prime\prime}$,
15$^{\prime\prime}$, 20$^{\prime\prime}$, 30$^{\prime\prime}$, and
40$^{\prime\prime}$.  The maximum value of the aperture,
40$^{\prime\prime}$, is set by the contrast of the image relative to
the background. The infrared colors of an annulus with internal and
external radii of 15$^{\prime\prime}$ and 30$^{\prime\prime}$,
respectively, centered on the peak of the emission in NGC~1068 are
also shown (filled circle).  
\\ 
\\
Fig. 5 -- The stellar continuum ratios log[F(0.16)/F(2.18)] (left panel) and
log[F(1.29)/F(2.18)] (right panel) are plotted as a function of the color
excess measured from H$\beta$/Br$\gamma$. The typical 1~$\sigma$ uncertainties
on the data are shown in  the lower left corner of the figures. The linear fit
to the data, discussed in the text, is also shown. The positions of NGC4861 and
IC1586 are marked in the left and right panels, respectively, and are not
included in the fits of the data. 
\\
\\
Fig. 6 -- The obscuration curve derived from the stellar continuum flux 
densities of
the starburst galaxies in the sample. The vertical axis shows the difference
k$^{\prime}$($\lambda$)$-$k$^{\prime}$(2.2), where the K-band (2.2~$\mu$m)
flux densities are the selected reference values. The horizontal axis is $\log
\lambda$($\mu$m). The filled circles are from the E(B$-$V)$_{H\alpha/H\beta}$
fits; the filled triangles are from the E(B$-$V)$_{H\beta/Br\gamma}$ fits. The
1 $\sigma$ uncertainties are reported. The obscuration curve derived by CKS94
is overlapped, with the V-band value arbitrarily chosen to be
k$^{\prime}$(0.55)$-$k$^{\prime}$(2.2)=2 (continuous line). 
\\
\\
Fig. 7 -- The observed flux density ratios at UV, optical and IR for
the 19 starburst galaxies (filled dots). The typical 1~$\sigma$
uncertainty on the data is shown at the bottom-right corner of the
plot. The two curved lines represent synthetic flux density ratios as
derived from models of stellar populations, as a function of the age
of the population. The arrows on the curves indicate the direction of
increasing age. The dotted curve shows the colors of an aging
population, originated from an instantaneous burst of star formation,
in the age range 1$\times$10$^7$--5$\times$10$^8$~yr. The dot-dashed
curve show the colors of a stellar population with constant star
formation rate in the age range
1$\times$10$^7$--1.5$\times$10$^{10}$~yr. The straight lines represent
the reddening trends, as derived from the CKS94 obscuration curve
(continuous lines), from the Galactic extinction (dashed line), and
from the LMC extinction (long-short dashed line). The reddening
increases from left to right. The intrinsic colors,
[F$_{\circ}$(0.55)/F$_{\circ}$(0.16);F$_{\circ}$(2.2)/F$_{\circ}$(0.16)],
are derived from the Template Starburst Spectrum (see Figure~13 and
section~6).  The positions of NGC~4861 and IC~214 are indicated.
\\
\\
Fig. 8 -- The color excess E(B$-$V) derived from the hydrogen Balmer/Brackett
line ratio, H$\beta$/Br$\gamma$, is shown as a function of the color excess
from the Balmer line ratio, H$\alpha$/H$\beta$. The error bars are 1 $\sigma$
uncertainties. Models of foreground clumpy dust are overplotted to the data; in
the scattering model, the radiation along the line of sight includes scattered
light from sources outside the observational beam (dashed line); in the
non-scattering model, the scattered light does not contribute to the observed
radiation (continuous line). The models are shown for an average of ${\cal
N}$=10 and $\cal N$=1000 clumps along the line of sight. 
\\
\\
Fig. 9 -- The selective obscuration produced by a foreground distribution of
dust clumps with: (1) $\tau_c(0.55)=0.37$ and $f={\cal N}^s/{\cal N}^g=1$,
using both the Galactic (1G) and the LMC (1L) extinction curves; (2)
$\tau_c(0.55)=0.37$ and $f=0.6$ (2G and 2L); (3) $\tau_c(0.55)=0.17$ and
$f=0.6$ (3G and 3L).  The parameter $f=0.6$ implies that the stellar continuum
radiation crosses 60\% of the dust clumps crossed by the nebular radiation. The
value of the clump optical depth $\tau_c(0.55)=0.37$ corresponds to
E(B$-$V)$_{H\alpha/H\beta}$=0.90 and E(B$-$V)$_{H\beta/Br\gamma}$=1.0;
$\tau_c(0.55)=0.17$ corresponds to E(B$-$V)$_{H\alpha/H\beta}$=0.51 and
E(B$-$V)$_{H\beta/Br\gamma}$=0.54, for ${\cal N}^g$=10.
\\
\\
Fig. 10 -- The H$_2$~v=1--0~S(1)(2.212~$\mu$m)/Br$\gamma$(2.166~$\mu$m)
ratio as a function of [FeII](1.257~$\mu$m)/Pa$\beta$(1.282~$\mu$m). The
positions of the LINER galaxy NGC~6764 and of the two starbursts NGC~4861 and
NGC~1569 are shown. Corrections for the dust reddening are negligible, and were
not applied. 
\\
\\
Fig. 11 -- The H$_2$~v=1--0~S(1)(2.212~$\mu$m) line flux as a function of the
Br$\gamma$(2.166~$\mu$m) flux for the starburst galaxies. The lines are
corrected for the effects of both foreground and intrinsic reddening using the
values of the color excess from column 4 of Table~1, and from the
H$\beta$/Br$\gamma$ ratios (see Table~4), respectively . The 1~$\sigma$ 
uncertainties are reported. 
\\
\\
Fig. 12 -- As Figure~11, now for the [FeII](1.257~$\mu$m) and the 
Pa$\beta$(1.282~$\mu$m) line fluxes.
\\
\\
Fig. 13 -- The template starburst spectrum (TSS), in units of flux
density as a function of wavelength, in the range
0.125--2.2~$\mu$m. The spectrum is normalized to the K-band flux
density, F(2.2). The error bar at the bottom left corner of each panel
gives a measure of the typical 1~$\sigma$ uncertainty. The spectrum
covers continuously the range 0.125-1.0~$\mu$m (solid line); the two
filled circles are the flux density values in the J and H bands. In
each panel, the template is overlapped to synthetic spectra (dotted
lines) produced from the Bruzual \& Charlot's (1996) and Leitherer \&
Heckman (1995) models, with Salpeter IMF in the range
0.1--100~M$_{\odot}$. The synthetic spectra are given by: a) a
2$\times$10$^7$~yr population with constant SFR; b) the combination of
two populations with constant SFR, the first 2$\times$10$^7$~yr old,
and the second 1.5$\times$10$^{10}$~yr old; c) a 2$\times$10$^9$~yr
population with constant SFR; d) a composite model made of a constant
SFR, 2$\times$10$^7$~yr old population, and two underlying
populations, 1$\times$10$^8$~yr and 5$\times$10$^8$~yr old, and both
generated by an instantaneous burst of star formation; the
contribution of each component is also shown, going from the blue
20~Myr population (dashed line), to the 100~Myr population (dot-dashed
line), to the red 500~Myr population (long-dashed line).
\\
\\
Fig. 14 -- The TSS (solid line $+$ filled circles) is overplotted to
the synthetic spectrum of a 5$\times$10$^7$~yr population with
constant SFR, and a Salpter IMF in the mass range 0.1--30~M$_{\odot}$
(dotted line). The scale and the axis labels are the same as in
Figure~13.  
\\ 
\\
Fig. 15 -- The spectrum of NGC~7714 (solid line for the UV and optical
continuum $+$ filled circles for the J and H band flux densities),
dereddened using the recipe of section~4.1, is overlapped to the best
fitting synthetic spectrum (dotted line), given by the combination of
a 1$\times$10$^7$~yr population with constant SFR, plus two underlying
populations, 1$\times$10$^8$~yr and 5$\times$10$^8$~yr old,
respectively, and originated from an instantaneous burst of star
formation (see text). The axis labels are as in Figure~13.
\\
\\
Fig. 16 -- The spectrum of NGC~4861 (solid line for the UV and optical
continuum $+$ filled circles for the J and H band flux densities),
dereddened using the recipe of section~4.1, is overplotted to best
fitting synthetic spectra (dotted lines): a) a 4$\times$10$^6$~yr,
constant SFR population with a 1.2$\times$10$^{10}$~yr burst
population; b) two burst populations, one 3$\times$10$^6$~yr old and
the other 1.2$\times$10$^{10}$~yr old. c) The spectrum of NGC~4861 is
dereddened using two different methods: the recipe of section~4.1
(solid line) and the Galactic extinction curve (dotted line). The axis
labels are as in Figure~13.
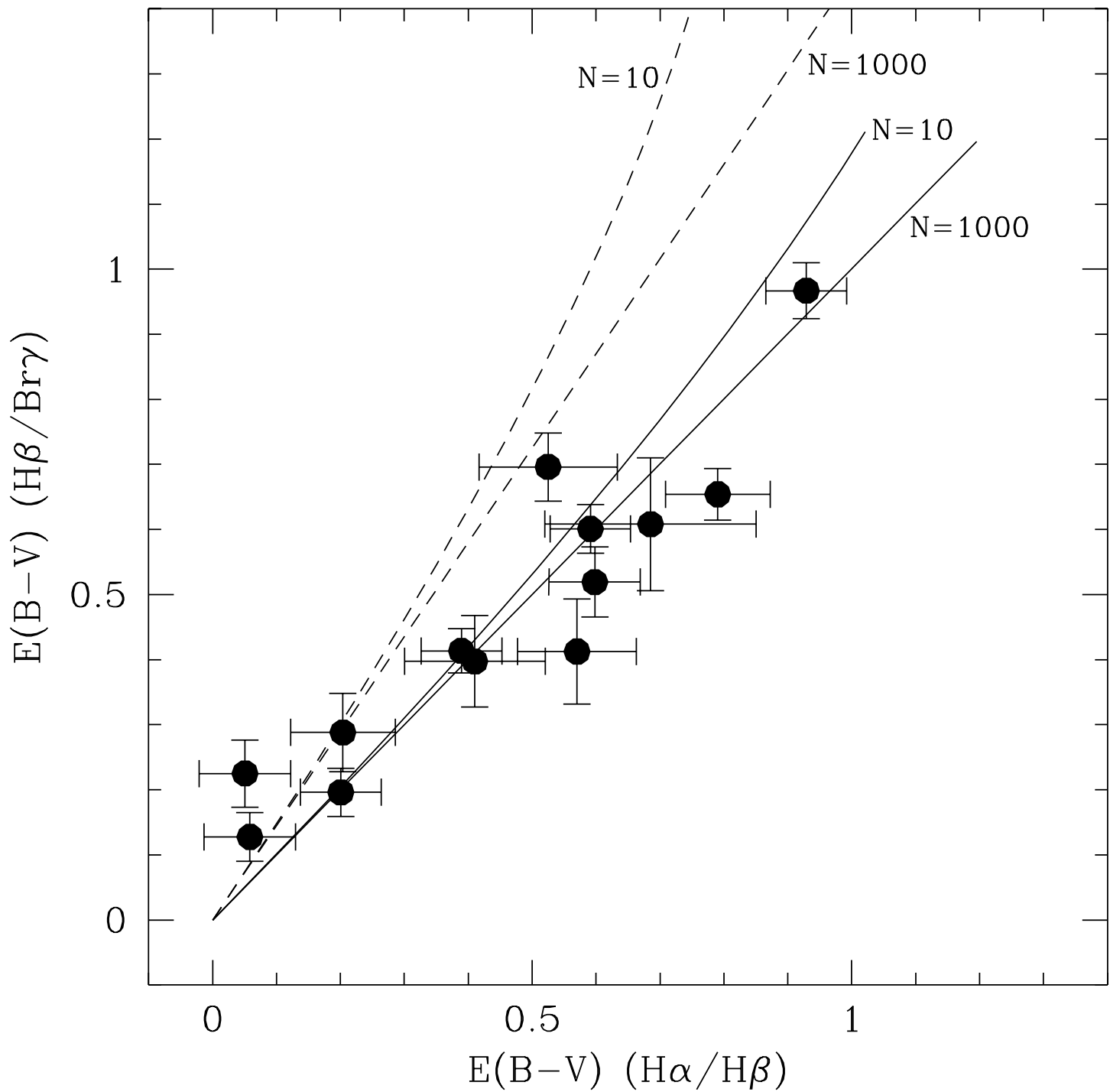
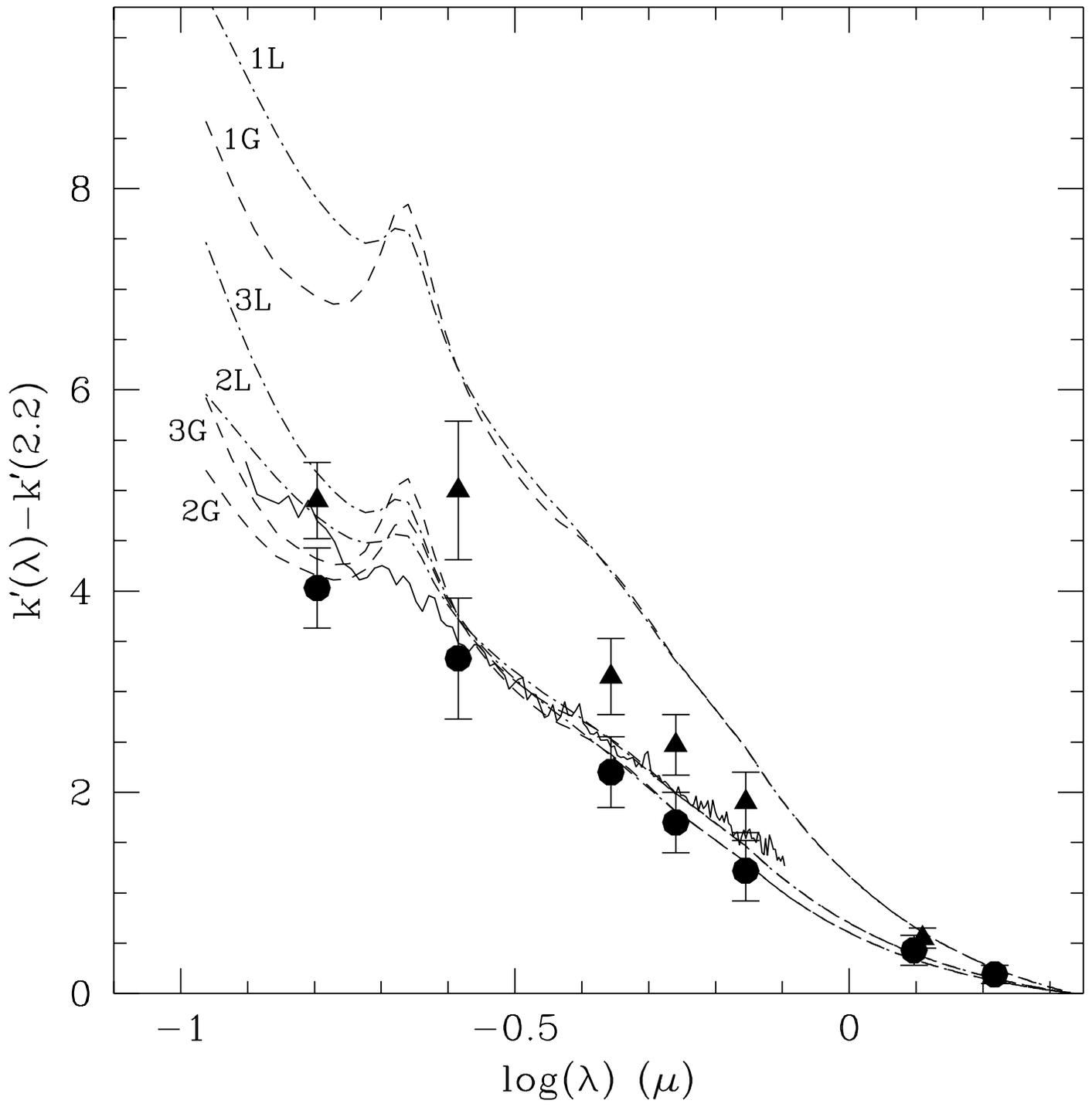
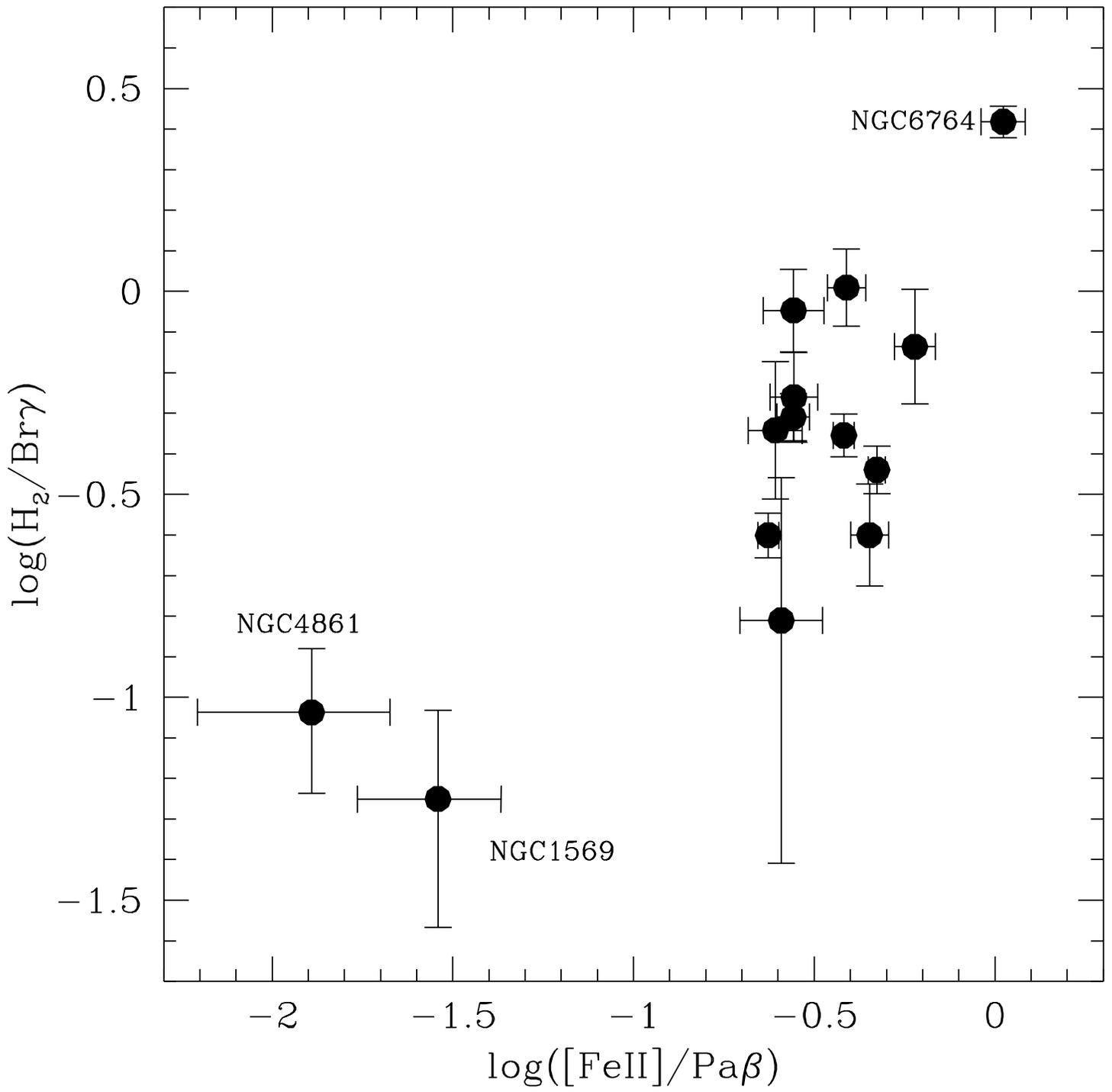
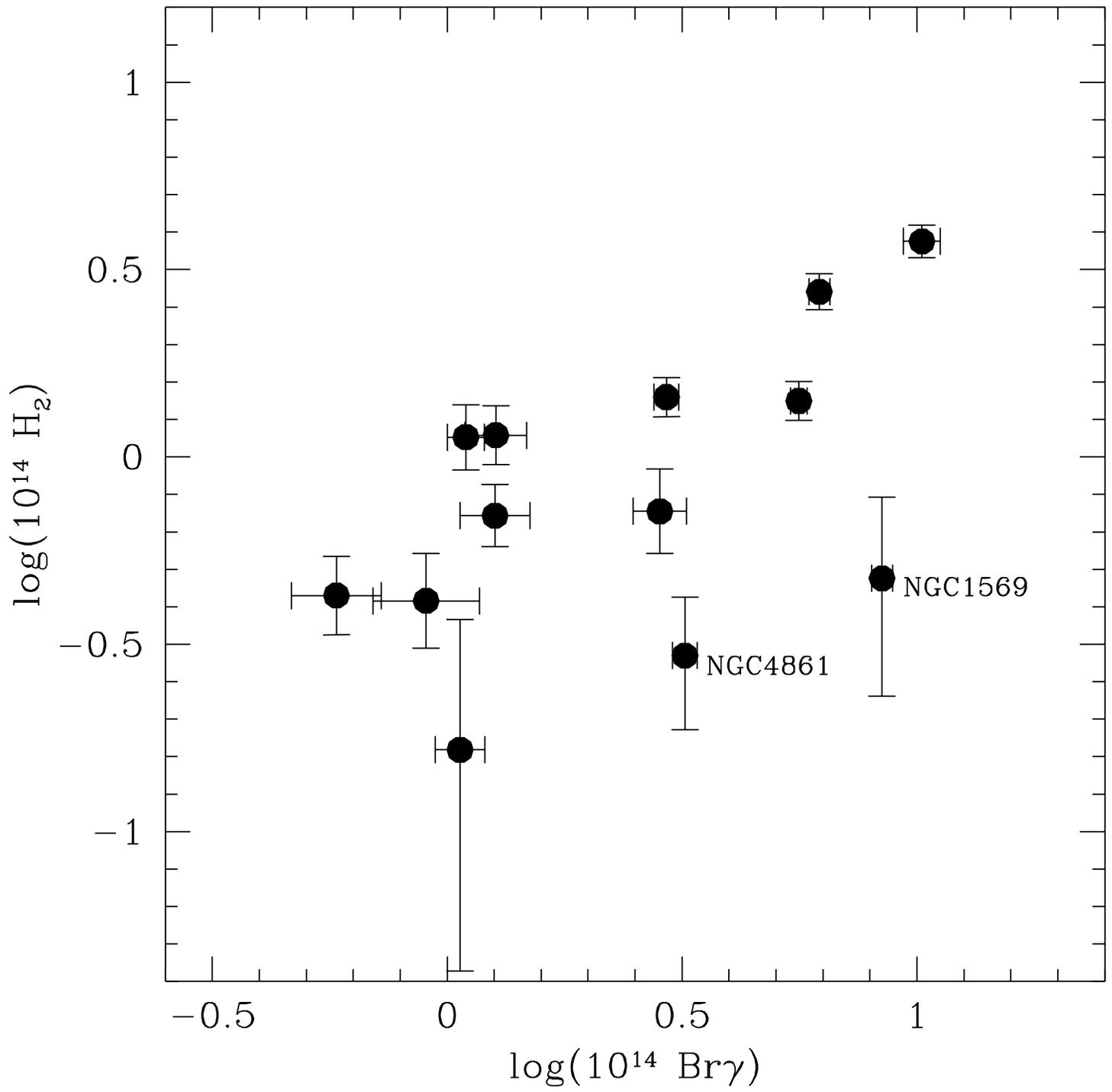
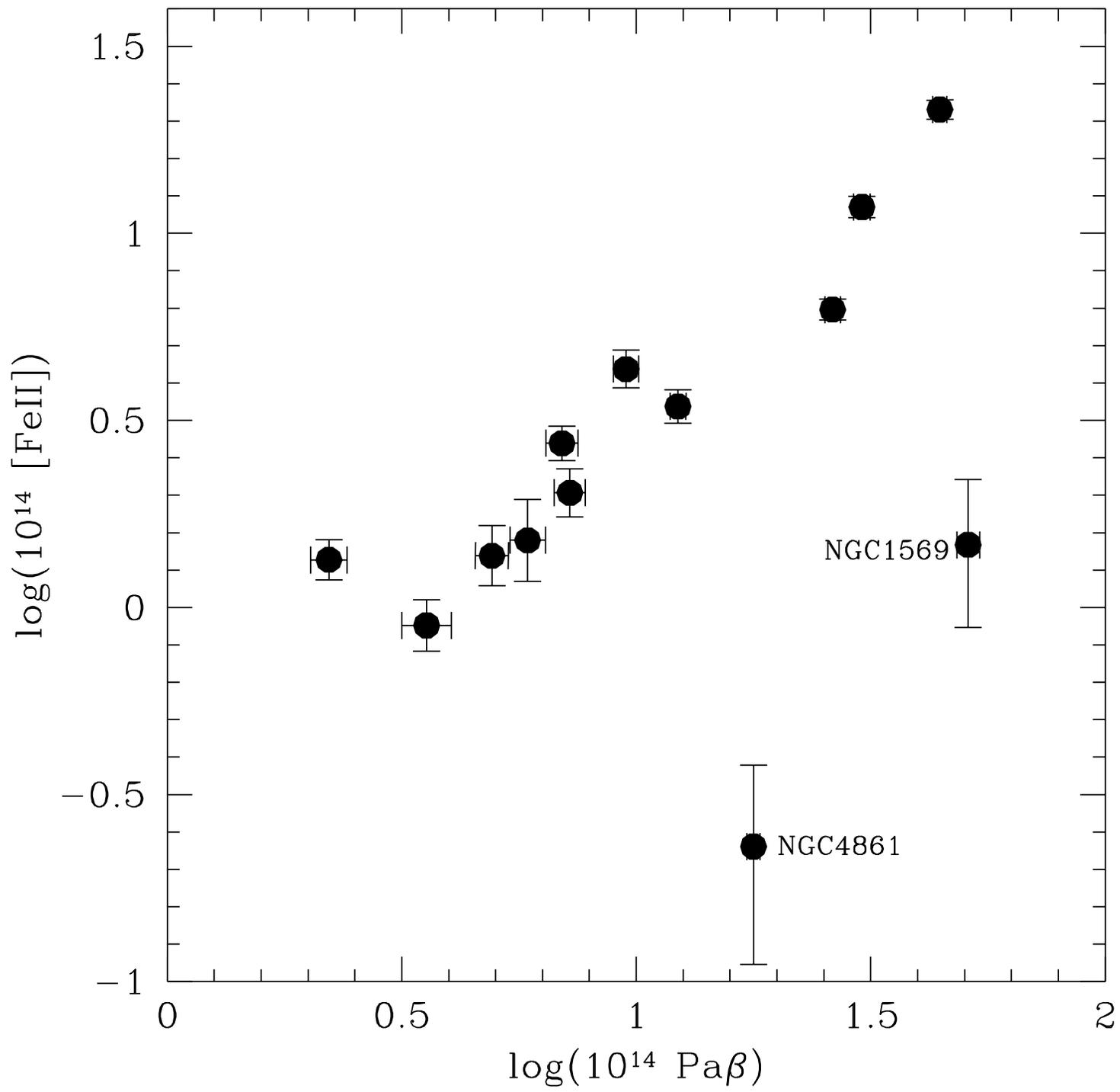
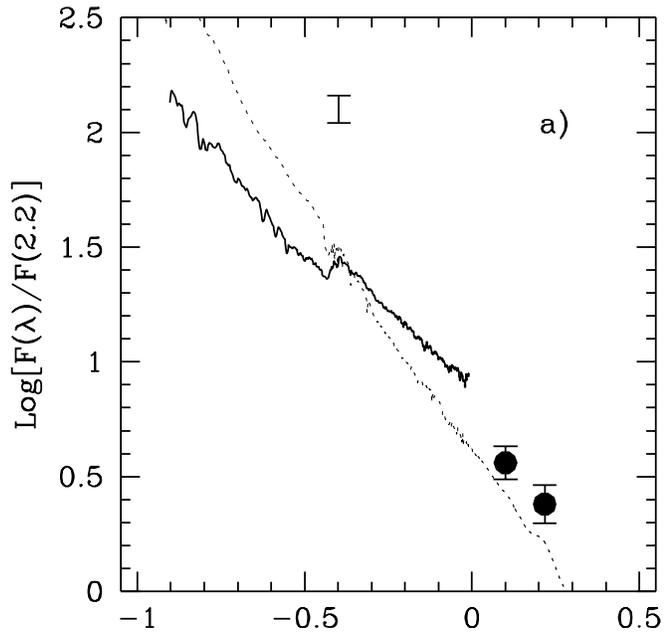
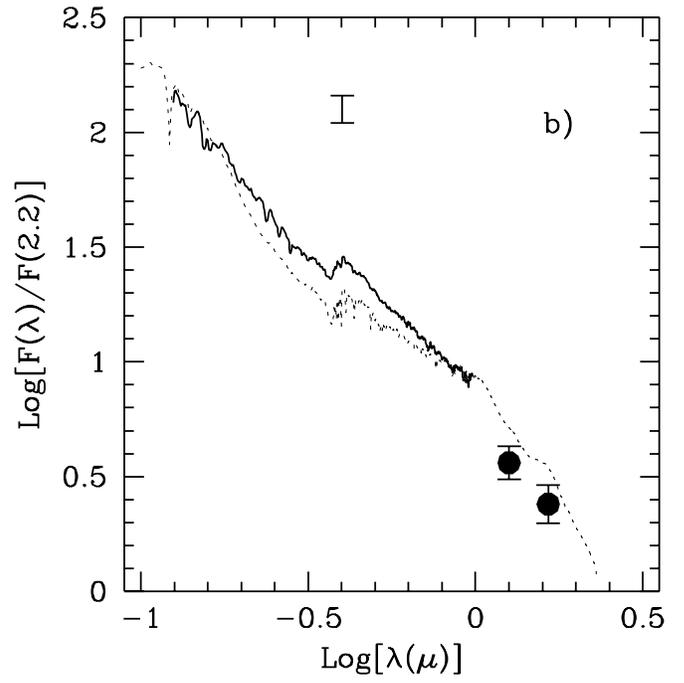
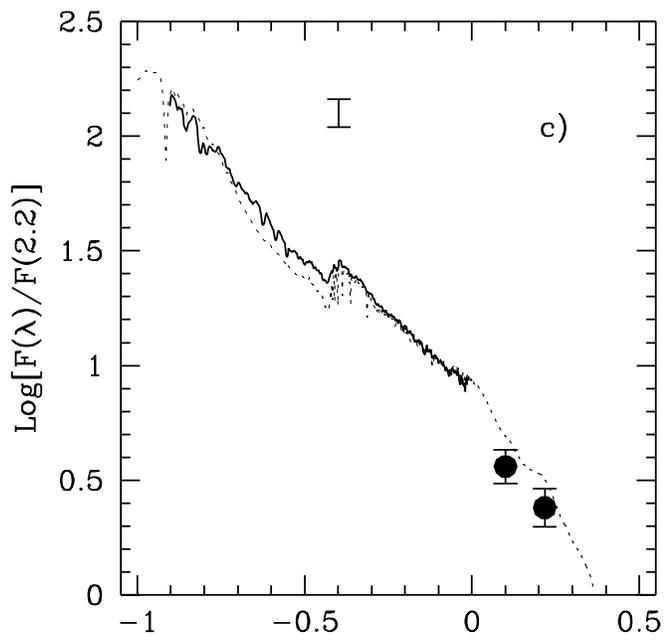
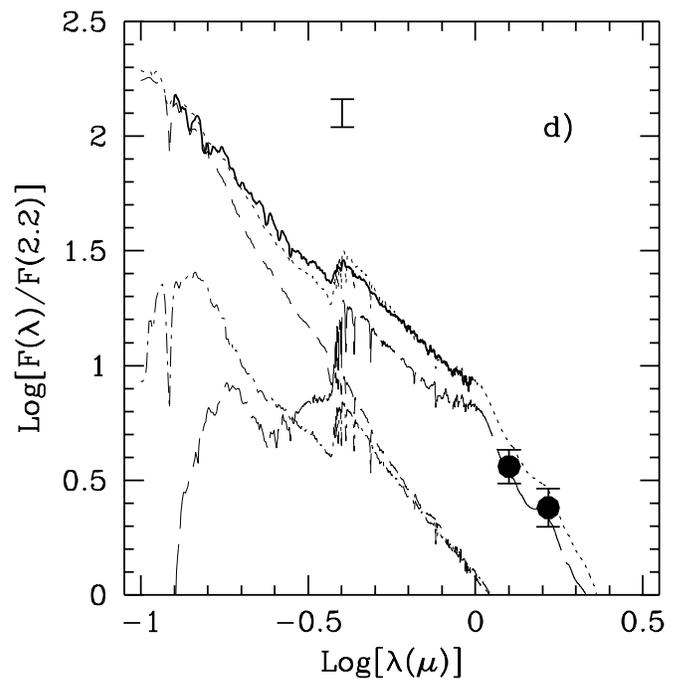
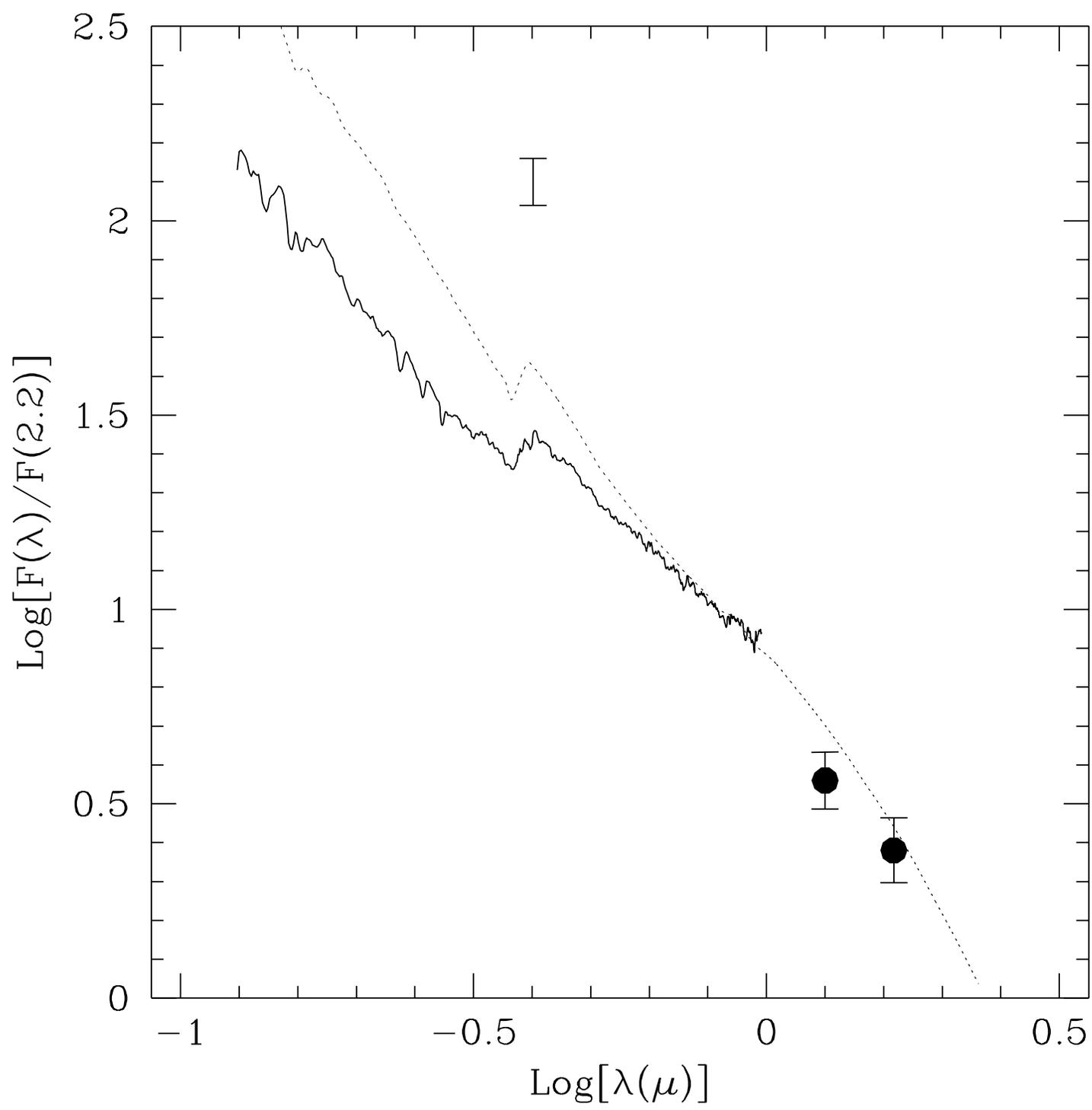
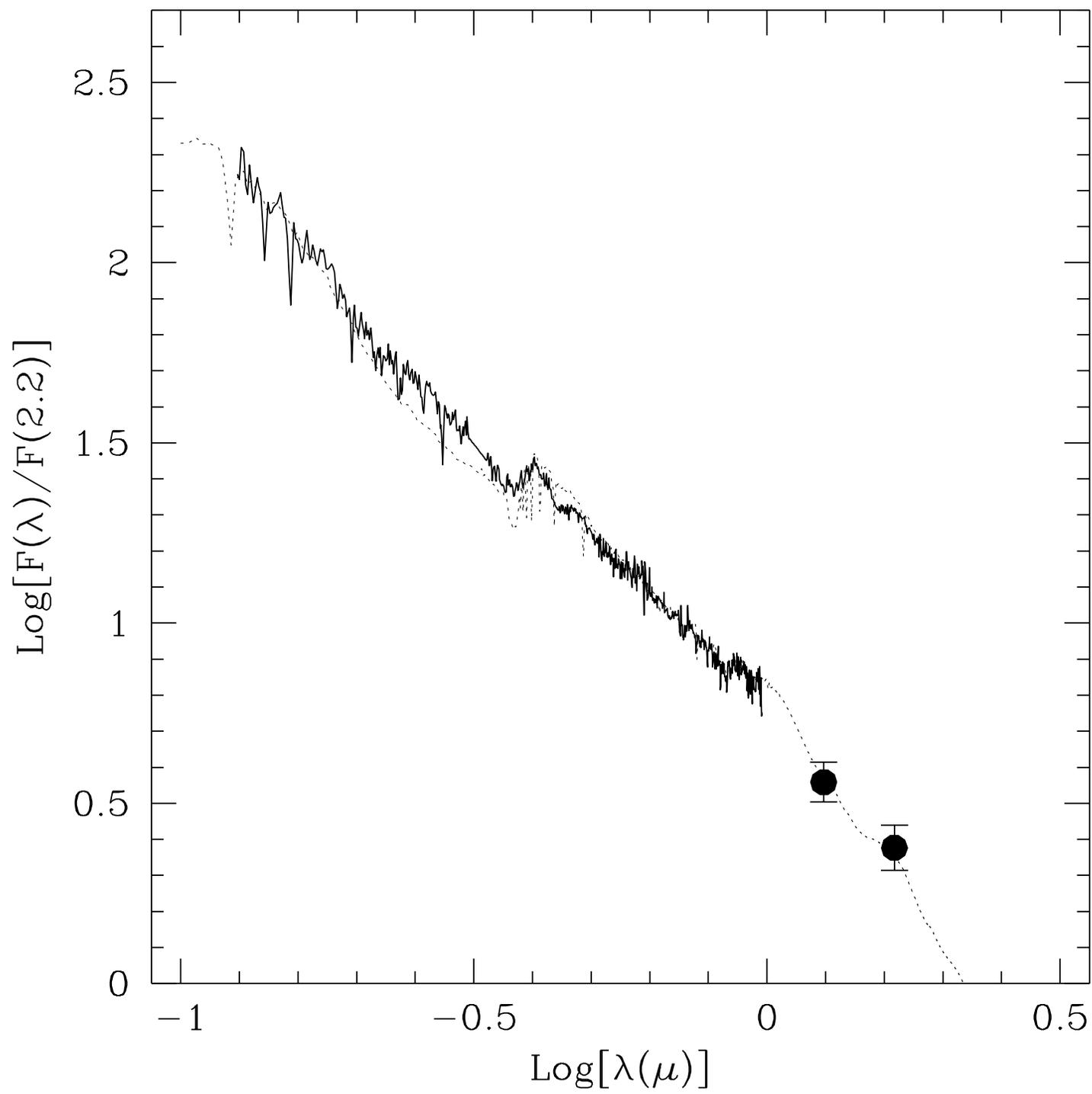
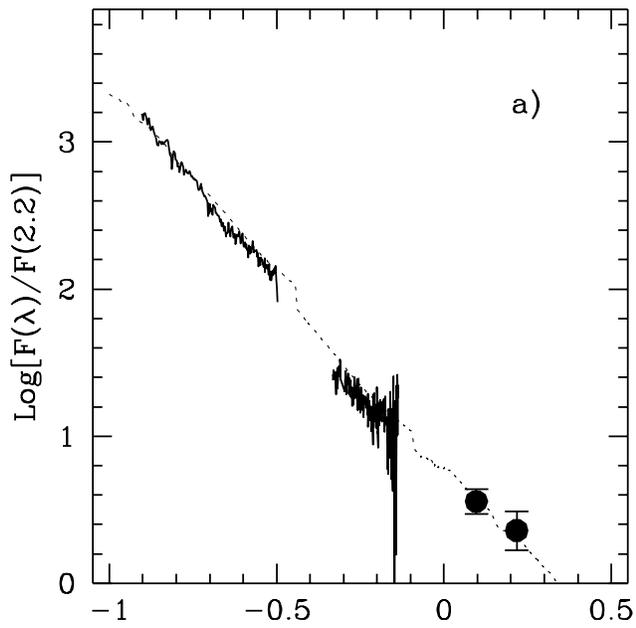
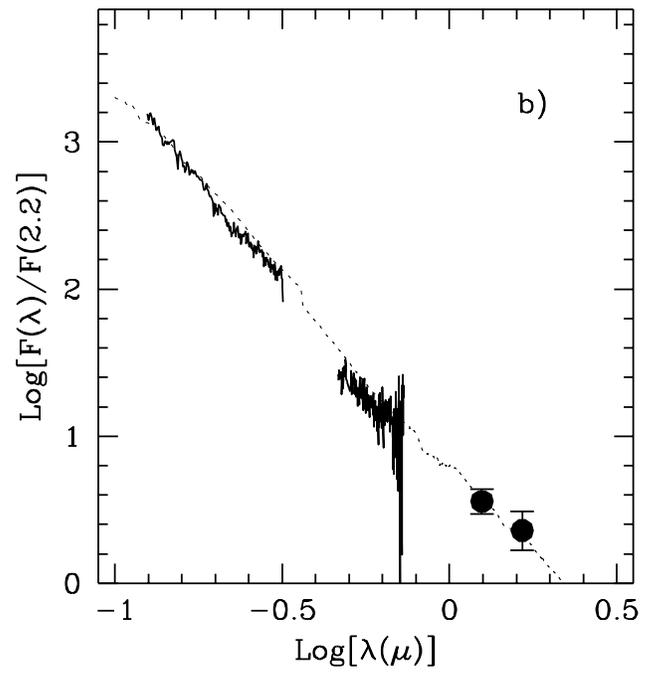
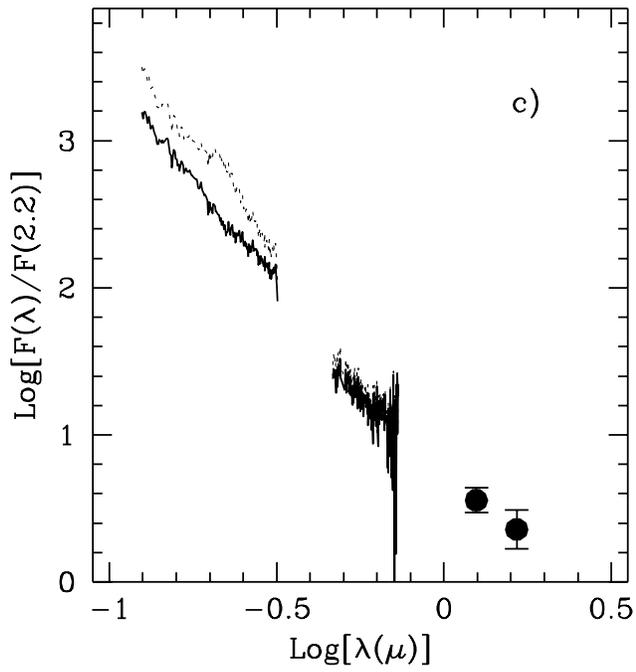
\end{document}